\documentclass[
  aps,prd,twocolumn,superscriptaddress,
  nofootinbib,floatfix,longbibliography
]{revtex4-2}

\usepackage{amsmath,amssymb}
\DeclareUnicodeCharacter{03C0}{\ensuremath{\pi}}
\DeclareUnicodeCharacter{03C3}{\ensuremath{\sigma}}
\usepackage{braket}
\usepackage{dsfont}

\usepackage{graphicx}
\usepackage{multirow}
\usepackage[normalem]{ulem}
\usepackage{xcolor}
\usepackage[colorlinks=true,linkcolor=blue,urlcolor=blue,citecolor=blue,anchorcolor=blue]{hyperref}
\usepackage[capitalize]{cleveref}
\Crefname{section}{Sec.}{Secs.}
\crefrangelabelformat{equation}{(#3#1#4--#5#2#6)}

\graphicspath{{fig/}}

\begin{document}

\title{Excited-State Effects in Nucleon Sigma Terms from a Variational Analysis}

\newcommand{\UCY}{Department of Physics,
University of Cyprus, P.O. Box 20537, 1678 Nicosia, Cyprus}
\newcommand{\CYI}{Computation-based Science and Technology Research Center,
The Cyprus Institute, 20 Kavafi Str., Nicosia 2121, Cyprus}

\author{Constantia Alexandrou} \affiliation{\UCY}\affiliation{\CYI}
\author{Giannis Koutsou} \affiliation{\CYI}
\author{Yan Li} \affiliation{\CYI}
\author{Srijit Paul} \affiliation{\UCY}
\author{Ferenc Pittler} \affiliation{\CYI}

\begin{abstract}
We investigate excited-state effects in nucleon sigma-terms in lattice
QCD using twisted-mass clover-improved fermion ensembles. We study
correlation functions with nucleon and multihadron
nucleon-$\sigma$-meson interpolating operators. The analysis is
performed using three gauge ensembles, including one ensemble
simulated with two degenerate light, a strange, and a charm quark with
quark masses tuned to approximately their physical value. We solve the
generalized eigenvalue problem that arises from the matrix of
correlators and show that the three-point function for the light
sigma-term obtained using the optimal linear combination of
interpolating operators has substantially reduced time dependence
without loss of statistical precision. We apply Laplace filtering and
investigate residual excited-state contributions in the extraction of
$\sigma_{\pi N}$. Using the same optimal linear combination of
correlators as for $\sigma_{\pi N}$ for the strange sigma-term we also
observe suppressed excited state contributions over an extended range
of source--sink time separations.  For the charm sigma-term, we
observe negligible time dependence within the large statistical
uncertainties, and these improvements show no significant effect.  %We
%obtain $\sigma_{\pi N}=45.7(3.0)(3.3)$~MeV,
%$\sigma_s=42.1(6.5)(2.3)$~MeV, and $\sigma_c=107.4(30.4)(0.6)$~MeV at
%the lattice spacing $a=0.07948(11)$~fm, where the first and second
%errors are statistical and systematic, respectively.
\end{abstract}

\maketitle

\section{Introduction}
\label{sec:introduction}
Nucleon sigma-terms quantify the scalar quark content of the nucleon and the explicit quark mass contribution to its mass. They provide essential hadronic input for Higgs-mediated interactions and scalar, spin-independent dark-matter scattering~\cite{Shifman:1978zn,Hoferichter:2017olk}. The $\pi N$ sigma term, $\sigma_{\pi N}$, characterizes explicit chiral symmetry breaking and can be determined phenomenologically from low-energy pion--nucleon scattering using Roy--Steiner equations and the Cheng--Dashen low-energy theorem~\cite{Cheng:1970mx,Hoferichter:2015dsa,RuizdeElvira:2017stg,Hoferichter:2023ptl}. The latest pionic-atom analysis gives $59.0(3.5)$~MeV, corresponding to $55.9(3.5)$~MeV in the isospin-symmetric QCD (isoQCD) convention adopted here~\cite{Hoferichter:2023ptl}, while several lattice calculations obtain values in the range $40$--$45$~MeV~\cite{RQCD:2022xux,Agadjanov:2023efe,Alexandrou:2024ozj}. The origin of this discrepancy remains under investigation~\cite{Alarcon:2021dlz}.

Lattice calculations determine sigma-terms either through the
Feynman--Hellmann theorem, using the quark-mass dependence of the
nucleon
mass~\cite{Durr:2015dna,Borsanyi:2020bpd,RQCD:2022xux,Liang:2025adz},
or directly from scalar matrix elements extracted from three-point
correlation
functions~\cite{Yang:2015uis,Gupta:2021ahb,Agadjanov:2023efe,Alexandrou:2024ozj,Park:2025rxi}. The
former requires controlled interpolation and extrapolation of the
nucleon mass near the physical point, while the latter is sensitive to
excited-state contamination. In particular, the scalar matrix elements
of low-lying multihadron states can contribute strongly to three-point
correlation functions while remaining suppressed in nucleon
two-point functions. Recent studies have reported shifts in
$\sigma_{\pi N}$ when such states are explicitly included in the
analysis~\cite{Gupta:2021ahb,Agadjanov:2023efe,Park:2025rxi},
highlighting the need for improved control of excited-state effects.

A complementary strategy is to enlarge the interpolator basis and solve a generalized eigenvalue problem (GEVP)~\cite{Luscher:1990ck,Blossier:2009kd}. Recent applications to nucleon scalar matrix elements have employed bases containing nucleon and multihadron interpolators~\cite{Alexandrou:2024tin,Barca:2024hrl,Wang:2025nsd,Barca:2026bmo}. The coupling of the scalar current to multihadron states can enhance their contributions to three-point functions~\cite{Barca:2025det}. In particular, the inclusion of $N\sigma$-type interpolators has been shown to reduce excited-state contamination in ensembles with $m_\pi=429$ and $222$~MeV, where the $\sigma$ is respectively stable and unstable~\cite{Barca:2024hrl,Barca:2026bmo}. Their effectiveness and the attainable precision near the physical pion mass, however, remain to be established.

The strange sigma-term, $\sigma_s$, is of particular interest because its uncertainty affects predictions for supersymmetric dark matter scattering~\cite{Ellis:2008hf}. Proposed extractions from near-threshold $J/\psi$ photoproduction involve substantial model dependence~\cite{Kou:2023zko}. Lattice calculations employ either direct matrix elements or the Feynman--Hellmann (F-H) method~\cite{Yang:2015uis,Agadjanov:2023efe,Alexandrou:2024ozj,Park:2025rxi,Barca:2024hrl,Borsanyi:2020bpd,Copeland:2021qni} although the latter yields larger statistical errors due to the mild dependence of the nucleon mass on the strange quark mass. The charm-sigma term, $\sigma_c$, provides a test of the heavy-quark expansion relating heavy-quark scalar operators to gluonic operators~\cite{Shifman:1978zn}; lattice results have found consistency between direct and F-H determinations and this expansion~\cite{Borsanyi:2020bpd,Alexandrou:2024ozj}.

In this work, we investigate whether a basis of nucleon and $N\sigma$
interpolators improves the extraction of the light, strange, and charm
sigma-terms using an ensemble of twisted-mass fermions with degenerate
light, strange, and charm quarks ($N_f=2+1+1$) with masses that
reproduce approximately the physical pion mass (referred to as physical point). We
compare standard and GEVP-improved three-point functions for the sigma-terms, assessing excited-state contamination and statistical precision
for each quark flavor. We further employ Laplace
filtering~\cite{Portelli:2025lop} to probe residual excited-state
contributions and study excited states using two additional ensembles
with different pion masses, volumes, and dynamical quark-flavor
content. The remainder of this paper is structured as follows: The
simulation setup and scalar operators are described in
\cref{sec:simulation-details}; the two-point analysis and variational
and Laplace constructions are presented in
\cref{sec:two-point-analysis}; the light and the strange and charm matrix
elements are discussed in
\cref{sec:three-point-analysis,sec:strange-charm}, respectively; and
our conclusions are given in \cref{sec:conclusions}.

\section{Simulation Details}
\label{sec:simulation-details}

We use three gauge ensembles generated by the Extended Twisted Mass Collaboration (ETMC) with twisted-mass clover-improved fermions. Two are physical point ensembles, while the third is simulated at a heavier pion mass. The simulations are performed at maximal twist, which ensures automatic ${\cal O}(a)$ improvement of physical observables~\cite{Frezzotti:2000nk,Frezzotti:2003ni}. The ensemble parameters are summarized in \cref{tab:ensemble}.
\begin{table*}
\caption{Parameters of the gauge ensembles used in this work. The
  ensemble name is provided in the first column, with its abbreviation
  in parenthesis.  In the following columns we provide the number of
  sea-quark flavors ($N_f$), the lattice volume $(L/a)^3\times T/a$ in
  terms of the spatial ($L$) and temporal ($T$) extents, the lattice
  spacing ($a$), the spatial extent ($L$), the pion mass ($m_\pi$),
  the dimensionless product of the pion mass and spatial extent
  ($m_\pi L$), and the number of gauge configurations analyzed
  ($N_{\rm cfg}$). The lattice spacings and pion masses are taken from
Refs.~\cite{ExtendedTwistedMass:2022jpw,ExtendedTwistedMass:2024nyi,Alexandrou:2018sjm,Alexandrou:2018egz,ETM:2015ned}.}
\label{tab:ensemble}
\centering
\begin{ruledtabular}
\begin{tabular}{cccccccc}
Ensemble & $N_f$ & $(L/a)^3\!\times T/a$ & $a$ [fm] & $L$ [fm] & $m_\pi$ [MeV] & $m_\pi L$ & $N_{\rm cfg}$ \\ \hline
\texttt{cA211.53.24} (A24) & $2+1+1$ & $24^3\!\times48$ & 0.09076(54) & 2.18 & 362(2) & 3.98 & 2467 \\
\texttt{cA2.09.48} (A48) & $2$ & $48^3\!\times96$ & 0.0938(3) & 4.50 & 130.6(4) & 2.98 & 1228 \\
\texttt{cB211.072.64} (B64) & $2+1+1$ & $64^3\!\times128$ & 0.07948(11) & 5.09 & 140.2(2) & 3.62 & 732
\end{tabular}
\end{ruledtabular}
\end{table*}
We focus on the physical-point ensemble \texttt{cB211.072.64} (B64),
with $N_f=2+1+1$ dynamical quark flavors. For comparison, we also
analyze the heavier-pion ensemble \texttt{cA211.53.24} (A24), with the
same sea-quark content, and the physical-point ensemble
\texttt{cA2.09.48} (A48), with $N_f=2$ degenerate light sea
quarks. These complementary results are presented in
Appendix~\ref{app:a24-a48}. Details regarding the generation of the
ensembles and the determination of the lattice spacing ($a$) are given
in
Refs.~\cite{ETM:2015ned,Alexandrou:2018egz,ExtendedTwistedMass:2022jpw,ExtendedTwistedMass:2024nyi}.

In our setup, we  use nucleon interpolating operators with the
nucleon at rest and nucleon-$\sigma$-meson interpolating operators
where both the nucleon and $\sigma$-meson are at rest. The standard
proton interpolator ${\cal J}_N$ and the scalar-isoscalar interpolator
${\cal J}_\sigma$ and their product ${\cal J}_{N\sigma}$ in this setup are given by,
\begin{align}
{\cal J}_N(t)&=\sum_{\vec{x}}\epsilon_{abc}\big[u^a(\vec{x},t)^T{\cal C}\gamma_5d^b(\vec{x},t)\big]u^c(\vec{x},t), \label{eq:nucleon-interpolator}\\
{\cal J}_\sigma(t)&=\frac{1}{\sqrt{2}}\sum_{\vec{x}}\big[\bar u^a u^a+\bar d^a d^a\big](\vec{x},t), \label{eq:sigma-interpolator}\\
{\cal J}_{N\sigma}(t)&={\cal J}_N(t){\cal J}_\sigma(t). \label{eq:Nsigma-interpolator}
\end{align}
Here ${\cal C}$ is the charge-conjugation matrix and $a,b,c$ are color
indices, with repeated indices summed. Dirac indices are suppressed,
including the open Dirac index of ${\cal J}_N(t)$ and ${\cal
  J}_{N\sigma}$. The bilinear in ${\cal J}_\sigma$ has $I=0$ and
$J^P=0^+$. Thus, $\sigma$ labels the scalar-isoscalar channel, to which
several finite-volume states can contribute.

We apply Gaussian smearing~\cite{Gusken:1989ad,Alexandrou:1992ti} to
all quark fields entering the nucleon and sigma interpolators using
\begin{align}
\psi^a_{\rm smear}(\vec{x},t)&=\sum_{\vec{y}}F^{ab}(\vec{x},\vec{y};t)\psi^b(\vec{y},t), \label{eq:smearing}\\
F&=(\mathds{1}+\alpha_GH)^{N_G}, \label{eq:smearing-kernel}\\
H(\vec{x},\vec{y};t)&=\sum_{i=1}^3\left[U_i(\vec{x},t)\delta_{\vec{x},\vec{y}-\hat{i}}+U_i^\dagger(\vec{x}-\hat{i},t)\delta_{\vec{x},\vec{y}+\hat{i}}\right]. \label{eq:hopping-matrix}
\end{align}
For the B64 ensemble, we use $N_G=125$ and $\alpha_G=0.2$. For the A24 and A48 ensembles, we use $N_G=50$ and $\alpha_G=4.0$, as in Ref.~\cite{Alexandrou:2024tin}. For all three ensembles, the spatial links entering $H$ are APE smeared~\cite{APE:1987ehd} with $n_{\rm APE}=50$ and $\alpha_{\rm APE}=0.5$. The smearing parameters correspond to a nucleon root-mean-square radius of approximately $0.5$~fm.
\begin{figure}[!h]
\centering
\includegraphics[width=\columnwidth]{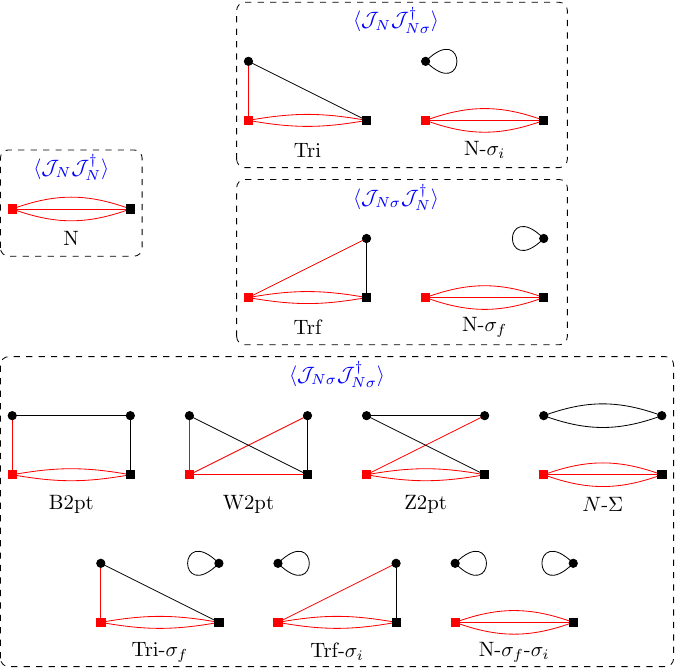}
\caption{Quark-flow topologies for the two-point correlation matrix $C^{\rm 2pt}_{jk}$, with $j,k\in\{N,N\sigma\}$. Squares and circles denote nucleon and scalar-isoscalar interpolators, respectively, with the source on the left and sink on the right. Red squares indicate the nucleon source, and red lines denote point-source propagators. Each group is labeled by its correlator, with time arguments suppressed. Hyphens in the topology names separate quark-disconnected pieces. Further computational details are given in Ref.~\cite{Alexandrou:2024tin}.}
\label{fig:diags-2pt}
\end{figure}

\begin{figure}[!h]
\centering
\includegraphics[width=\columnwidth]{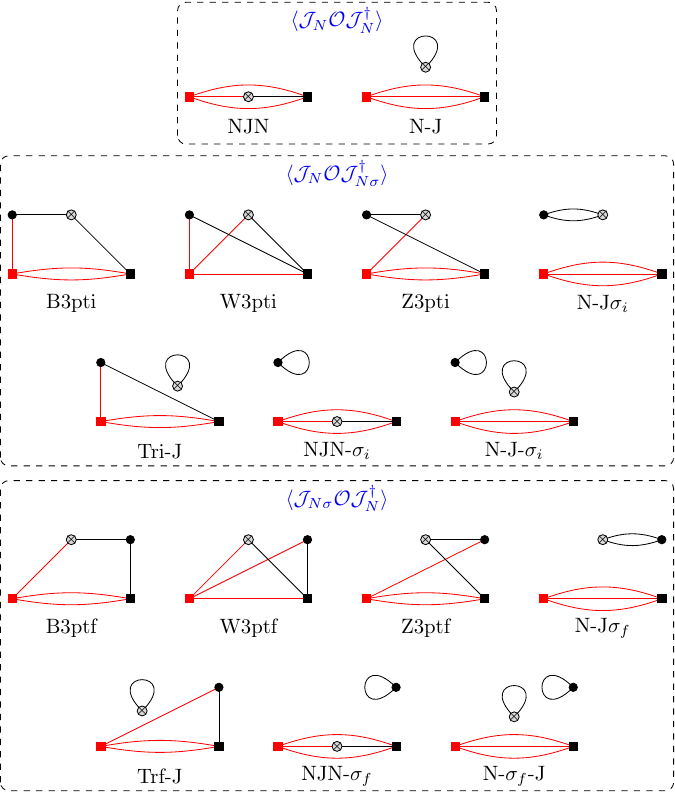}
\caption{Quark-flow topologies for the three-point functions $\braket{{\cal J}_N\mathcal O{\cal J}_N^\dagger}$, $\braket{{\cal J}_N\mathcal O{\cal J}_{N\sigma}^\dagger}$, and $\braket{{\cal J}_{N\sigma}\mathcal O{\cal J}_N^\dagger}$. A crossed circle denotes the scalar insertion. The remaining notation is as in \cref{fig:diags-2pt}.}
\label{fig:diags-3pt}
\end{figure}

\begin{figure}[!h]
\centering
\includegraphics[width=\columnwidth]{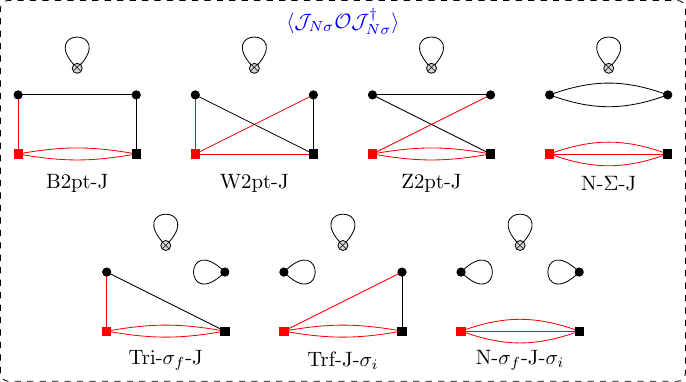}
\caption{Quark-flow topologies for $\braket{{\cal J}_{N\sigma}\mathcal O{\cal J}_{N\sigma}^\dagger}$ with a strange- or charm-quark insertion, using the notation of \cref{fig:diags-2pt,fig:diags-3pt}.}
\label{fig:diags-3pt-full}
\end{figure}

The resulting zero-momentum two-point correlation matrix is
\begin{align}
C^{\rm 2pt}_{jk}(t)=\mathrm{Tr}{[\Gamma_0\braket{{\cal J}_j(t){\cal J}_k^\dagger(0)}]},\qquad j,k\in\{N,N\sigma\},
\label{eq:two-point-matrix}
\end{align}
where the projector $\Gamma_0=\frac{1+\gamma_0}{2}$ projects to
positive parity and the trace is over Dirac indices.  We use the
mass-weighted scalar insertion operators
\begin{align}
\mathcal O\in\left\{m_\ell(\bar u u+\bar d d),\ m_s\bar s s,\ m_c\bar c c\right\}.
\label{eq:scalar-insertions}
\end{align}
Here $m_\ell$ is the degenerate light-quark mass, and $m_s$ and $m_c$ are the strange- and charm-quark masses. In the twisted-mass formulation, the quark-mass and scalar-density renormalization factors cancel in these combinations, so no renormalization factor is required for the sigma-terms. The matrix element $\braket{N|\mathcal O|N}$ at zero momentum transfer gives $\sigma_{\pi N}$, $\sigma_s$, or $\sigma_c$, respectively. With the insertion summed over its spatial position, at source--sink separation $t_s$ and insertion time $t_{\rm ins}$ we compute
\begin{align}
C^{\rm 3pt}_{jk}(t_s,t_{\rm ins})=\mathrm{Tr}{[\Gamma_0\braket{{\cal J}_j(t_s)\mathcal O(t_{\rm ins}){\cal J}_k^\dagger(0)}]},
\label{eq:three-point-matrix}
\end{align}
for $(j,k)=(N,N),(N,N\sigma),(N\sigma,N)$. The light-quark three-point correlator
contains connected and disconnected contributions, while the strange-
and charm-quark three-point correlators are purely disconnected. The
light-current entry with $(j,k)=(N\sigma,N\sigma)$ is not computed because
its connected contribution is computationally demanding. We instead use
the weighted construction of Ref.~\cite{Alexandrou:2024tin}, which cancels
ground-to-excited-state transitions in the two-state limit without
requiring this entry, as described in \cref{sec:standard-three-point}.
The quark-flow topologies
for the two- and three-point functions are shown in
\cref{fig:diags-2pt,fig:diags-3pt} respectively. For the strange and
charm currents, which are disconnected, the case
$(j,k)=(N\sigma,N\sigma)$ can be constructed from the
existing two-point contractions and quark loops without additional
propagator inversions and we
therefore include it, completing the $2\times2$ three-point
matrix. The seven topologies, which include either a strange- or
charm-quark loop attached to a two-point topology through gauge
fields, are shown in \cref{fig:diags-3pt-full}. For the computation of
the topologies in
\cref{fig:diags-2pt,fig:diags-3pt,fig:diags-3pt-full}, we use
combinations of sequential inversions and stochastic methods for
sampling the all-to-all propagators, following the $N\pi$ calculation
described in Ref.~\cite{Alexandrou:2024tin}. 

In
\cref{tab:statistics-sources}, we provide details of the statistics
available in terms of the number of source-positions per configuration
and the sink-source separations for each diagram class. In
\cref{tab:statistics-projected}, we provide the number of stochastic
vectors used for the all-to-all propagator estimation for each diagram
class. For the B64 ensemble in particular, we deflate 200 eigenvectors
for the calculation of the light-quark loops ($N_{\rm def}$=200) and
use fully color and spin diluted stochastic sources ($N_{\rm dil}$=12)
partitioned in space and time using hierarchical probing to distance
8, which results in 512 Hadamard vectors ($N_{\rm Had}$=512). For the
strange and charm quark loops we do not deflate and use $N_{\rm
  vec}$=2 color and spin diluted vectors with hierarchical probing to
distance 8 for the strange and $N_{\rm vec}$=12 color and spin diluted
vectors with hierarchical probing to distance 4 ($N_{\rm Had}$=32) for
the charm.

\begin{table}[!h]
\caption{Source-position statistics for each ensemble and contraction class. The last two columns give the source--sink separations $t_s/a$ and the number of source positions per gauge configuration $N_{\rm src}$. The topology labels refer to \cref{fig:diags-2pt,fig:diags-3pt,fig:diags-3pt-full}, with $X=B,W,Z$.}
\label{tab:statistics-sources}
\centering
\scriptsize
\begin{ruledtabular}
\begin{tabular}{c c c c}
Ensemble & Subset & $t_s/a$ & $N_{\rm src}$ \\ \hline
\multirow{4}{*}{A24}
 & $N$ & all $t_s$ & $121$ \\
 & $NJN$ & $10,12,14$ & $16$ \\
 & $T,X_{\rm 2pt}$ & all $t_s$ & $12$ \\
 & $X_{\rm 3pt}$ & $10,12,14$ & $12$ \\ \hline
\multirow{4}{*}{A48}
 & $N$ & all $t_s$ & $272$ \\
 & $NJN$ & \begin{tabular}{@{}c@{}}$10,12,14$\\$16,18$\end{tabular} & \begin{tabular}{@{}c@{}}$9$\\$32$\end{tabular} \\
 & $T,X_{\rm 2pt}$ & all $t_s$ & $4$ \\
 & $X_{\rm 3pt}$ & $10,12,14,16,18$ & $4$ \\ \hline
\multirow{4}{*}{B64}
 & $N$ & all $t_s$ & $369$ \\
 & $NJN$ & \begin{tabular}{@{}c@{}}$8,10,12,14$\\$16,18,20$\end{tabular} & \begin{tabular}{@{}c@{}}$1,2,5,10$\\$32,112,128$\end{tabular} \\
 & $T,X_{\rm 2pt}$ & all $t_s$ & $20$ \\
 & $X_{\rm 3pt}$ & \begin{tabular}{@{}c@{}}$8,10,12,14$\\$16,18,20$\end{tabular} & \begin{tabular}{@{}c@{}}$4$\\$20$\end{tabular}
\end{tabular}
\end{ruledtabular}
\end{table}

\begin{table}[!h]
\caption{Number of stochastic vectors used for the fully spatially
  projected measurements, organized by contraction class and
  ensemble. We give the number of stochastic time-slice sources
  ($N_{\rm stoc}$) and one-end-trick sources ($N_{\rm oet}$). For the
  quark loops of the B64 ensemble, we provide the number of
  eigenvectors deflated ($N_{\rm def}$), and the total number of
  vectors inverted ($n_{\rm inv}$) in terms of the product $n_{\rm inv}=N_{\rm
    dil}\times N_{\rm Had}\times N_{\rm vec}$, as explained in the
  text. The contraction labels are those of
  \cref{tab:statistics-sources}. Dashes indicate measurements not
  included in this study.}
\label{tab:statistics-projected}
\centering
\scriptsize
\begin{ruledtabular}
\begin{tabular}{c c c c}
Subset & A24 & A48 & B64 \\ \hline
$\sigma_{i,f}$ & $N_{\rm stoc}=200$ & $N_{\rm stoc}=100$ & $N_{\rm stoc}=25$ \\
$B,W$ & $N_{\rm stoc}=12$ & $N_{\rm stoc}=12$ & $N_{\rm stoc}=12$ \\
$\Sigma,J\sigma_{i,f},Z$ & $N_{\rm oet}=1$ & $N_{\rm oet}=1$ & $N_{\rm oet}=1$ \\
$J$ (light) & $N_{\rm stoc}=200$ & $N_{\rm stoc}=400$ & \begin{tabular}{@{}c@{}}$N_{\rm def}=200$\\$n_{\rm inv}=12\!\times\!512\!\times\!1$\end{tabular} \\
$J$ (strange) & -- & -- & \begin{tabular}{@{}c@{}}$N_{\rm def}=0$\\$n_{\rm inv}=12\!\times\!512\!\times\!2$\end{tabular} \\
$J$ (charm) & -- & -- & \begin{tabular}{@{}c@{}}$N_{\rm def}=0$\\$n_{\rm inv}=12\!\times\!32\!\times\!12$\end{tabular}
\end{tabular}
\end{ruledtabular}
\end{table}

Both ${\cal J}_\sigma$ and $\mathcal O$ have nonzero vacuum expectation values. We therefore use the vacuum-subtracted operators
\begin{align}
{\cal J}_\sigma\to{\cal J}_\sigma-\braket{{\cal J}_\sigma},\qquad
\mathcal O\to\mathcal O-\braket{\mathcal O}, \label{eq:vacuum-subtraction}
\end{align}
with vacuum subtraction understood throughout.

\subsection{Pion-Mass Corrections to the Sigma Terms}
\label{sec:pion-mass-corrections}

For making quantitative comparisons near the physical point, we correct for the small
mistuning of the pion masses for the A48 and B64 ensembles relative to the
isospin-symmetric QCD (isoQCD) value $m_\pi^\star=135.0$~MeV. At
leading order in chiral perturbation theory (ChPT), the pion-mass
dependence of $\sigma_{\pi N}=m_\ell g_S^\ell$ follows from
$m_\pi^2\propto m_\ell$ and the mass independence of the scalar charge
$g_S^\ell=\braket{N|\bar uu+\bar
  dd|N}$~\cite{Hoferichter:2015hva}. For the B64 ensemble, we thus use
the values of $am_\ell^{\rm iso}=0.0006669(28)$, $am_s^{\rm
  iso}=0.018267(53)$, and $am_c^{\rm iso}=0.23134(52)$ in
\cref{eq:scalar-insertions}, taken from
Ref.~\cite{ExtendedTwistedMass:2024nyi}, rather than the simulated
bare quark masses $am_\ell^{\rm sim}=0.00072$, $am_s^{\rm
  sim}=0.0186$, and $am_c^{\rm sim}=0.249$. For the A48 ensemble such an
isoQCD tuning has not been carried out. We thus use
\begin{align}
am_\ell^{\rm iso}=am_\ell^{\rm sim}\left(\frac{m_\pi^\star}{m_\pi^{\rm sim}}\right)^2,
\label{eq:light-mass-lo-correction}
\end{align}
which gives the isoQCD bare light-quark mass $am_\ell^{\rm iso}=0.000962(6)$ from the simulated value $am_\ell^{\rm sim}=0.0009$. Results for the heavier-pion A24 ensemble  are quoted at its simulated bare light-quark mass, $am_\ell^{\rm sim}=0.0053$. This leading prescription requires no ChPT low-energy constants. The higher-order changes to the pion-mass corrections for the A48 and B64 ensembles are estimated to be less than $1$~MeV, as shown in Appendix~\ref{app:pion-mass-corrections}.

The phenomenological determination $\sigma_{\pi N}^{\pi^\pm}=59.0(3.5)$~MeV uses the charged-pion mass to define the isospin limit~\cite{Hoferichter:2023ptl}. To obtain the value at the neutral-pion convention, following Ref.~\cite{Hoferichter:2023ptl}, one needs to apply  downward shifts of $3.7(1)$, $3.2(1)$, and $3.1(1)(4)$~MeV at $\mathcal O(p^2)$, $\mathcal O(p^3)$, and $\mathcal O(p^4)$, respectively. Adopting the $\mathcal O(p^4)$ correction gives
\begin{align}
\sigma_{\pi N}^{\rm iso}
&=\sigma_{\pi N}^{\pi^\pm}-3.1(1)(4)~\mathrm{MeV}
 =55.9(3.5)~\mathrm{MeV}.
\label{eq:sigma-phenomenology-isoqcd}
\end{align}

\section{Analysis of Two-Point Functions}
\label{sec:two-point-analysis}

Our analysis of the two-point functions includes multi-state fits, a
GEVP analysis, and the Laplace filter
method~\cite{Portelli:2025lop}. We briefly describe the three
approaches in the following, before presenting our results for all
three methods in the case of the two-point function.
\subsection{Multi-State Fits}
\label{sec:standard-two-point}

For $k\in\{N,N\sigma\}$, we define the overlap of the interpolating
operator ${\cal J}_k$ with the $n^{\rm th}$ eigenstate of the
Hamiltonian as $Z_{k,n}=\braket{\Omega|{\cal J}_k|n}$. The spectral
decomposition of the corresponding standard two-point function is
\begin{align}
C^{\rm 2pt}_{kk}(t)=\sum_n\left|Z_{k,n}\right|^2e^{-E_nt}. \label{eq:c2pt-spectral}
\end{align}
We construct the effective mass as
\begin{align}
a m_{{\rm eff},k}(t)=\log\frac{C^{\rm 2pt}_{kk}(t)}{C^{\rm 2pt}_{kk}(t+a)}. \label{eq:c2pt-meff}
\end{align}
We fit $m_{{\rm eff},N}$ using one-, two-, and three-state truncations of \cref{eq:c2pt-spectral}, denoting the ground-state energy by $E_0=m_N$ and the earliest timeslice included in the fit by $t_{\rm low}$.

\subsection{GEVP Analysis}
\label{sec:two-point-gevp}

To suppress excited-state contamination associated with the $N\sigma$ channel, we solve a $2\times2$ generalized eigenvalue problem (GEVP) 
\begin{align}
C^{\rm 2pt}(t)v_n(t,t_0)=\lambda_n(t,t_0)C^{\rm 2pt}(t_0)v_n(t,t_0), \label{eq:gevp}
\end{align}
using the $N$ and $N\sigma$ interpolators introduced in \cref{sec:simulation-details}.
In Eq.~(\ref{eq:gevp}), $t_0$ is the reference time, while $\lambda_n$ and $v_n$ are the
principal correlator and eigenvector for state $n$, respectively. The
eigenvectors approach their asymptotic limits when both $t$ and $t_0$
are sufficiently large~\cite{Luscher:1990ck,Blossier:2009kd}. We study
this limit at fixed $t-t_0$, following the $N\pi$ analysis in
Ref.~\cite{Alexandrou:2024tin}.  The components of the eigenvectors,
$v_{nk}$, with $k\in\{N,N\sigma\}$, specify the linear combination of
interpolators for channel $n$ and at zero momentum are
real-valued. We will refer to the linear combination,
\begin{align}
\widetilde{\cal J}_N^\dagger(t)=\sum_{k\in\{N,N\sigma\}}v_{0k}(t,t_0){\cal J}_k^\dagger(t), \label{eq:optimized-interpolator}
\end{align}
as the ``GEVP-improved'' nucleon interpolating operator.
\cref{eq:optimized-interpolator} suggests that the coefficients of the
linear combination depend on time, but in practice we fit
$v_{0,N\sigma}/v_{0,N}$ to a constant keeping the difference $t-t_0$
fixed to $t-t_0=2a$, thus fixing the relative weights between ${\cal
  J}_N$ and ${\cal J}_{N\sigma}$.  Using the GEVP-improved nucleon
interpolating operator, we form $C_{\rm GEVP}^{\rm
  2pt}(t)=\braket{\widetilde{\cal J}_N(t)\widetilde{\cal
    J}_N^\dagger(0)}$ and analyze this correlation function using
multi-state fits as described in Sec.~\ref{sec:standard-two-point} for
the standard correlator. We will denote energies extracted from
GEVP-improved correlators with a tilde.

Analogously, the GEVP-improved $N\sigma$ interpolator is
\begin{align}
\widetilde{\cal J}_{N\sigma}^\dagger(t)=\sum_{k\in\{N,N\sigma\}}v_{1k}(t,t_0){\cal J}_k^\dagger(t). \label{eq:optimized-Nsigma-interpolator}
\end{align}

\subsection{Laplace Filter Method}
\label{sec:two-point-laplace}

As an alternative treatment of excited-state contributions, we apply
the Laplace-filter technique used for lattice correlators in
Ref.~\cite{Portelli:2025lop}. In continuous Euclidean time, the filter
is defined by
\begin{align}
{\cal L}^{\rm cont}_E=-\partial_t^2+E^2. \label{eq:laplace-cont}
\end{align}
Acting on the spectral decomposition in \cref{eq:c2pt-spectral}, it gives
\begin{align}
{\cal L}^{\rm cont}_E C^{\rm 2pt}_{kk}(t)
=\sum_n (E^2-E_n^2)|Z_{k,n}|^2 e^{-E_nt}. \label{eq:laplace-cont-spectrum}
\end{align}
The filter leaves the energies unchanged and multiplies the spectral weight of state $n$ by $E^2-E_n^2$. Tuning $E$ to $E_n$ removes contributions at that energy.

The discretized symmetric second derivative with displacement $\delta$ gives
\begin{align}
{\cal L}_{E,\delta}C(t)&=\left[-\nabla_\delta^2+\lambda^2(E,\delta)\right]C(t),\nonumber\\
\nabla_\delta^2 C(t)&=\frac{C(t+\delta)+C(t-\delta)-2C(t)}{\delta^2},\nonumber\\
\lambda(E,\delta)&=\frac{\sqrt{e^{E\delta}+e^{-E\delta}-2}}{\delta}. \label{eq:laplace-filter}
\end{align}
In the limit $\delta\to0$, $\nabla_\delta^2\to\partial_t^2$ and $\lambda(E,\delta)\to E$, we recover \cref{eq:laplace-cont}. Acting on \cref{eq:c2pt-spectral}, the discrete filter gives
\begin{align}
{\cal L}_{E,\delta}C^{\rm 2pt}_{kk}(t)
=\sum_n\left[\lambda^2(E,\delta)-\lambda^2(E_n,\delta)\right]|Z_{k,n}|^2e^{-E_nt}. \label{eq:laplace-response}
\end{align}
The choice $E=E_1$ removes the  contribution of the first-excited state at any $\delta$. We apply this filter to the standard nucleon and GEVP-improved correlators. We will refer to the filtered effective mass as the effective mass of the filtered correlator and denote it with a superscript ${\rm Lap}$,
\begin{align}
a m_{{\rm eff},k}^{\rm Lap}(t;E)=\log\frac{{\cal L}_{E,\delta}C^{\rm 2pt}_{kk}(t)}{{\cal L}_{E,\delta}C^{\rm 2pt}_{kk}(t+a)}. \label{eq:laplace-meff}
\end{align}
Our strategy is to fit the filtered effective mass to a constant $m$
and allow $E$ to vary as a fit parameter. We thus  minimize,
\begin{align}
\chi^2(E,m)&=\sum_{t,t'\in{\cal W}}\left[a m_{{\rm eff},k}^{\rm Lap}(t;E)-am\right]\nonumber\\
&\quad\times\left[\operatorname{Cov}^{-1}(E)\right]_{tt'}\left[a m_{{\rm eff},k}^{\rm Lap}(t';E)-am\right], \label{eq:laplace-chi2}
\end{align}
where ${\cal W}$ is the fit window and $\operatorname{Cov}(E)$ is the covariance matrix of the filtered effective mass. Both the filtered correlator and the covariance are recomputed at each trial $E$. We denote the fitted nucleon mass and first-excited-state gap by $m_N^{\rm Lap}$ and $\Delta E_1^{\rm Lap}$. The filtered three-point ratios in the next section are analyzed in the same way.

The Laplace method does not determine the overlap ratio $|Z_{k,1}/Z_{k,0}|^2$ and therefore has one fewer fit parameter than the two-state effective-mass fit.

Smaller temporal displacements $\delta$ give noisier but less-correlated filtered data. The fitted energies are nevertheless only weakly dependent on the displacement, as shown in Appendix~\ref{app:laplace-discretization}.

\begin{figure*}[!t]
\centering
\includegraphics[width=\textwidth]{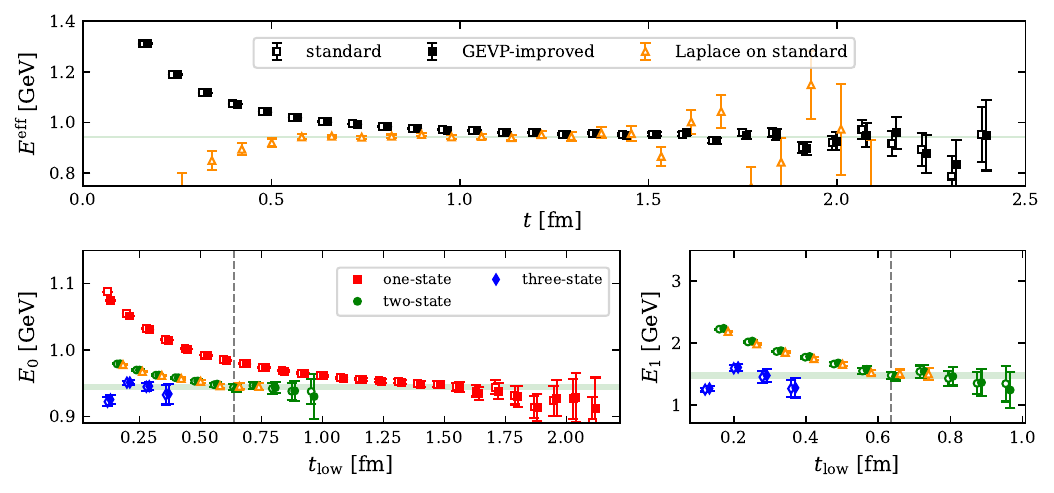}
\caption{Results on the ground state and excited states of the nucleon from the analysis of its effective mass using the standard and GEVP-improved nucleon interpolators. Top: effective masses from the standard correlator (open black squares), GEVP-improved correlator (filled black squares), and Laplace filter on the standard correlator (open orange triangles). Bottom: fitted $E_0$ (left) and $E_1$ (right) versus the lower fit bound $t_{\rm low}$. Red squares, green circles, and blue diamonds denote one-, two-, and three-state fits, respectively, with open and filled symbols distinguishing standard and GEVP-improved correlators. Orange triangles denote Laplace fits. Dashed lines mark the selected standard two-state fit, and green bands show its energies.}
\label{fig:c2pt-N-comparison}
\end{figure*}

\begin{figure*}[!t]
\centering
\includegraphics[width=\textwidth]{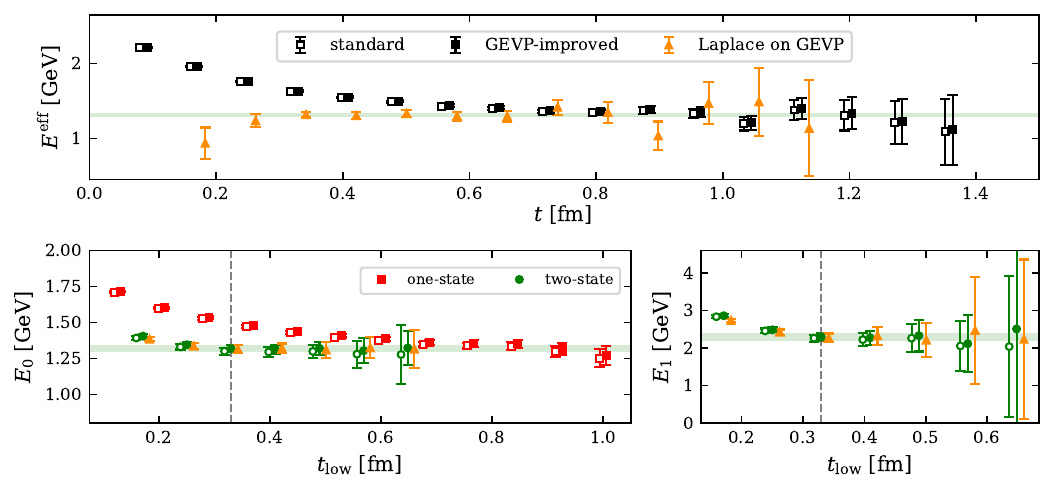}
\caption{Results on the lowest resolved state and excited  state  from the analysis of the effective mass using the $N\sigma$ two-point function. The notation is the same as that   of \cref{fig:c2pt-N-comparison}. Here the Laplace filter acts on the GEVP-improved correlator (filled orange triangles). Dashed lines and green bands identify the selected GEVP-improved two-state fit.}
\label{fig:c2pt-Nsigma-comparison}
\end{figure*}

\subsection{Results of two-point function analysis}
The results of our effective-mass analysis using multi-state fits, GEVP,
and the Laplace filter method are summarized in
Figs.~\ref{fig:c2pt-N-comparison} and~\ref{fig:c2pt-Nsigma-comparison}
for the nucleon and $N\sigma$, respectively. For the multi-state
analysis, one-, two-, and three-state fits to the standard and
GEVP-improved effective mass are carried out, while for the filtered
effective mass, we consider the procedure described in
Sec.~\ref{sec:two-point-laplace} as a two-state fit. As can be seen
from the lower left panel of \cref{fig:c2pt-N-comparison}, as $t_{\rm
  low}$ is increased, the three approaches converge to a common nucleon
mass.  

\begin{figure}[!htbp]
\centering
\includegraphics[width=\columnwidth]{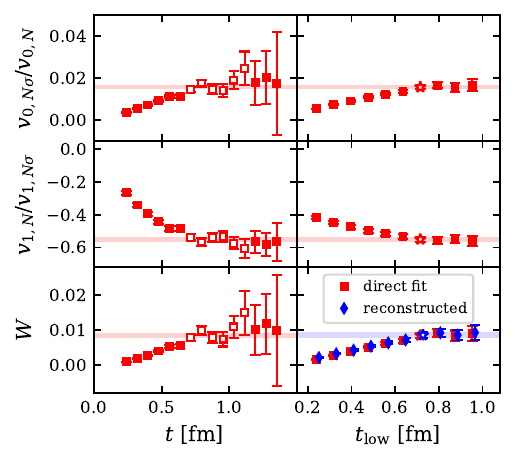}
\caption{GEVP eigenvector-component ratios and weight $W$. Rows show $v_{0,N\sigma}/v_{0,N}$, $v_{1,N}/v_{1,N\sigma}$, and $W$. Left: dependence on $t$ at fixed $t-t_0=2a$, with open squares identifying the fitted points. Right: constant fits versus $t_{\rm low}$, with open stars marking the selected fits. Red bands show the corresponding direct-fit results and uncertainties. In the bottom-right panel, red squares show direct fits to $W$, while blue diamonds and the blue band show its reconstruction from the component fits.}
\label{fig:gevp-vw}
\end{figure}

While the standard $N\sigma$ correlator has the same spectral
decomposition as the nucleon, its overlaps favor different states over
the accessible time range in~\cref{fig:c2pt-Nsigma-comparison}. As can
be seen, the one-state fits approach the two-state result as $t_{\rm
  low}$ is increased. In Fig.~\ref{fig:gevp-vw}, we show the
ratios $v_{0,N\sigma}/v_{0,N}$ and $v_{1,N}/v_{1,N\sigma}$ which we
fit to in order to construct the time-independent linear combinations
for the GEVP-improved interpolating operators (see
Sec.~\ref{sec:two-point-gevp}). The fitted value for
$v_{0,N\sigma}/v_{0,N}=0.01564(99)$ shown in the top row of
\cref{fig:gevp-vw} is rather small, suggesting that the nucleon
effective mass changes little after GEVP improvement, as confirmed in
\cref{fig:c2pt-N-comparison}. The two-operator basis, thus, does not
seem to substantially suppress excited-state contamination in this
correlator. The standard and GEVP-improved two-state fits yield
consistent energies at sufficiently late $t_{\rm low}$, while GEVP
improvement slightly increases the  uncertainties at larger times. We
therefore use the standard effective mass, adopting $t_{\rm
  low}/a=8$, the same window for the two-state fit adopted in
Ref.~\cite{Alexandrou:2026oks} using the same ensemble, and find
$m_N=944.0(4.4)$~MeV and $\Delta E_1^{\rm 2pt}=E_1^{\rm
  2pt}-m_N=531(55)$~MeV .

The same analysis for the coefficients is carried out for $n=1$,
giving $v_{1,N}/v_{1,N\sigma}=-0.550(17)$, as shown in the second row
of \cref{fig:gevp-vw}. As suggested by the value of the ratio, in this
case, GEVP improvement has a more significant effect. Although the
individual standard and GEVP-improved energy errors overlap in
\cref{fig:c2pt-Nsigma-comparison}, their correlated difference
resolves an increase of $17.5(2.3)$~MeV at $t_{\rm low}/a=4$. This
upward shift is consistent with removing a small contribution from the
ground-state nucleon, to which the standard $N\sigma$ interpolator
also couples.

Thus, for the $N\sigma$, we adopt the GEVP-improved two-state fit at $t_{\rm low}/a=4$, which gives $\widetilde E_0^{N\sigma}=1.317(25)$~GeV and $\widetilde E_1^{N\sigma}=2.29(10)$~GeV. We define $\Delta \widetilde E_0^{N\sigma}=\widetilde E_0^{N\sigma}-m_N$ and $\Delta \widetilde E_1^{N\sigma}=\widetilde E_1^{N\sigma}-\widetilde E_0^{N\sigma}$. The first compares the lowest resolved $N\sigma$ energy with the nucleon mass, while the second is the energy gap between $\widetilde E_1^{N\sigma}$ and $\widetilde E_0^{N\sigma}$. Here $\widetilde E_0^{N\sigma}$ is the lowest energy resolved over the fit window, not the ground state of the common nucleon channel.

We use the Laplace filtering method to obtain the energies shown in
Figs.~\ref{fig:c2pt-N-comparison} and~\ref{fig:c2pt-Nsigma-comparison}
illustrated with the orange triangles. Using $\delta/a=2$, we find that the
Laplace filtered results agree with the corresponding two-state fits
both in their central values and uncertainties. The filtered nucleon
effective mass approaches the plateau from below because the higher
excited states  acquire a spectral weight with the opposite sign.

\begin{figure}[!htbp]
\centering
\includegraphics[width=\columnwidth]{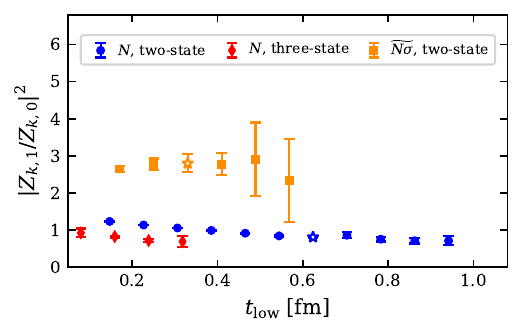}
\caption{Squared overlap ratios $|Z_{k,1}/Z_{k,0}|^2$ versus $t_{\rm low}$. Blue circles and red diamonds show standard nucleon two- and three-state fits, and orange squares show GEVP-improved $N\sigma$ two-state fits. Open stars replace the selected two-state points. GEVP-improved $N\sigma$ results with absolute uncertainty greater than two are omitted.}
\label{fig:c2pt-overlap-comparison}
\end{figure}

The three-point construction uses the normalization-independent weight
defined in Ref.~\cite{Alexandrou:2024tin},
\begin{align}
W(t,t_0)=\frac{1}{v_{0N}(t,t_0)[v^{-1}]_{N0}(t,t_0)}-1=-\frac{v_{0,N\sigma}}{v_{0,N}}\frac{v_{1,N}}{v_{1,N\sigma}}, \label{eq:W}
\end{align}
where $v$ denotes the matrix with entries $v_{nk}$ and all components are evaluated at $(t,t_0)$. The bottom row of \cref{fig:gevp-vw} compares a direct fit to $W$ with the product of the two separately fitted component ratios. The two determinations are nearly identical over the time range where convergence is observed. We use the reconstructed value of $W=0.00861(78)$.

The overlap ratios quantify the relative excited-state contributions within each truncation: the first excitation contributes $|Z_{k,1}/Z_{k,0}|^2e^{-(E_1-E_0)t}$ relative to the lowest resolved state. For the GEVP-improved interpolator, we define $Z_{\widetilde{N\sigma},n}=\braket{\Omega|\widetilde{\cal J}_{N\sigma}|n}$. The selected two-state fits in \cref{fig:c2pt-overlap-comparison} give $|Z_{N,1}/Z_{N,0}|^2=0.803(27)$ and $|Z_{\widetilde{N\sigma},1}/Z_{\widetilde{N\sigma},0}|^2=2.80(24)$. Thus, the GEVP-improved $N\sigma$ correlator has a larger relative excited-state overlap.

In a simple factorized picture, the two states would be $N\sigma$ and $N^*\sigma$, with $N^*$ a nucleon excitation and the same scalar state in both. If GEVP improvement only removes the ground-nucleon admixture, we would expect
\begin{equation*}
\left|\frac{Z_{\widetilde{N\sigma},1}}{Z_{\widetilde{N\sigma},0}}\right|^2
\simeq\left|\frac{Z_{N,1}}{Z_{N,0}}\right|^2.
\end{equation*}
The observed difference does not support this simple factorized interpretation. Since the $N\sigma$ channel couples to several interacting $N\pi\pi$-like levels, its fitted excitation can instead represent a combination of unresolved states.

\section{Analysis of \texorpdfstring{$\sigma_{\pi N}$}{sigma-pi-N}}
\label{sec:three-point-analysis}

The spectral decomposition of the scalar three-point function includes
transition matrix elements between the ground and excited states, such
as $\braket{0|\mathcal O|n}$ with $n>0$. These off-diagonal matrix
elements are absent from the two-point spectral decomposition and can
enhance contributions from states with small overlaps with the
interpolating field~\cite{Barca:2025det}. In the analysis of the
three-point function we therefore also explore fits where we allow the
corresponding energy gap to differ from that obtained in the two-point
function and examine the residual contamination after GEVP
improvement.

\subsection{Standard and GEVP-Improved Ratios for \texorpdfstring{$\sigma_{\pi N}$}{sigma-pi-N}}
\label{sec:standard-three-point}

The three-point function formed from ${\cal J}_N$ has the spectral decomposition
\begin{align}
C^{\rm 3pt}_{NN}(t_s,t_{\rm ins})=
\sum_{m,n}Z_{N,m}Z_{N,n}^*\braket{m|\mathcal O|n}\nonumber\\
\times e^{-E_m(t_s-t_{\rm ins})}e^{-E_nt_{\rm ins}}. \label{eq:c3pt-spectral}
\end{align}
The standard ratio is
\begin{align}
R_{\rm std}(t_s,t_{\rm ins})=\frac{C^{\rm 3pt}_{NN}(t_s,t_{\rm ins})}{C^{\rm 2pt}_{NN}(t_s)}, \label{eq:Rstd}
\end{align}
which approaches $\sigma_{\pi N}$ when both $t_{\rm ins}$ and $t_s-t_{\rm ins}$ become large. We symmetrize the ratio about $t_s/2$ and fit the independent insertion times $t_{\rm ins}\leq t_s/2$. A fit window is labeled by $(t_s^{\rm low}/a,t_{\rm ins}^{\rm cut}/a)$, where $t_s^{\rm low}$ is the smallest source--sink time separation and $t_{\rm ins}^{\rm cut}$ is the minimum distance of an included insertion from the source or sink.

\begin{figure*}[!t]
\centering
\includegraphics[width=\textwidth]{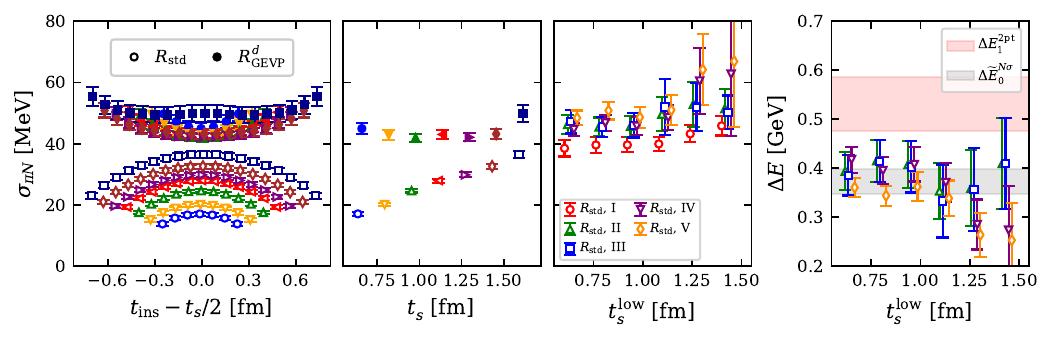}
\caption{Standard and GEVP-improved analyses of $\sigma_{\pi N}$. From
  left to right, in the first panel we show the insertion-time
  dependence of $R_{\rm std}$ (open) and $R_{\rm GEVP}^{d}$ (filled),
  in the second panel the ratios at the midpoint versus $t_s$, and in
  the third panel the fits to $R_{\rm std}$ using the five cases of
  \cref{eq:Rstd-two-state-cases}.  Colors and symbols identify the
  same source--sink separation in the first two panels. In the
  rightmost panel we show the values obtained for $\Delta E$ for the
  four of the five cases for which this is a fit parameter. The red and gray bands in the
  rightmost panel show $\Delta E_1^{\rm 2pt}$ and $\Delta\widetilde
  E_0^{N\sigma}$, respectively.}
\label{fig:Rstd-Rgevp}
\end{figure*}

We compare five fit forms based on the two-state spectral decomposition
\begin{subequations}\label{eq:Rstd-two-state-cases}
\begin{align}
R_{\rm I}&=\frac{g+r_{01}F(\Delta E_1^{\rm 2pt})+r_{11}e^{-\Delta E_1^{\rm 2pt}t_s}}
{1+r_1e^{-\Delta E_1^{\rm 2pt}t_s}}, \label{eq:Rstd-two-state-I}\\
R_{\rm II}&=\frac{g+r_{01}F(\Delta E)+r_{11}e^{-\Delta E t_s}}
{1+r_1e^{-\Delta E_1^{\rm 2pt}t_s}}, \label{eq:Rstd-two-state-II}\\
R_{\rm III}&=\frac{g+r_{01}F(\Delta E)+r_{11}e^{-\Delta E_1^{\rm 2pt}t_s}}
{1+r_1e^{-\Delta E_1^{\rm 2pt}t_s}}, \label{eq:Rstd-two-state-III}\\
R_{\rm IV}&=\frac{g+r_{01}F(\Delta E)}
{1+r_1e^{-\Delta E_1^{\rm 2pt}t_s}}, \label{eq:Rstd-two-state-IV}\\
R_{\rm V}&=g+r_{01}F(\Delta E), \label{eq:Rstd-two-state-V}
\end{align}
\end{subequations}
where $r_1=|Z_{N,1}/Z_{N,0}|^2$ and $\Delta E_1^{\rm 2pt}$ are fixed
from the two-point analysis, $g$ denotes the value of the sigma-term,
and we will refer to $r_{01}$ and $r_{11}$ as the transition and
diagonal amplitudes, associated with $\braket{0|\mathcal O|1}$ and
$\braket{1|\mathcal O|1}$, respectively.  The index $1$ labels the
effective excited state contribution retained in each term; the
contributions described by $r_1$, $r_{01}$, and $r_{11}$ need not
correspond to the same physical state. In
Eq.~(\ref{eq:Rstd-two-state-cases}), for the energy exponentials
associated with $\braket{0|\mathcal O|1}$ we use
\begin{align}
F(\Delta E;t_s,t_{\rm ins})=e^{-\Delta E t_{\rm ins}}+e^{-\Delta E(t_s-t_{\rm ins})}\label{eq:Rstd-transition}
\end{align}
with the arguments $t_s$ and $t_{\rm ins}$ suppressed in
\cref{eq:Rstd-two-state-cases}. Both exponentials remain after
symmetrization because $F(\Delta E;t_s,t_{\rm ins})$ is already
symmetric under $t_{\rm ins}\leftrightarrow t_s-t_{\rm ins}$, so
averaging the reflected data does not change its form. The five fit
forms of \cref{eq:Rstd-two-state-cases} differ in terms of the order
to which the spectral decomposition in the numerator and denominator
are truncated. They also differ in how the first excited state energy
is treated, namely whether it is fixed to $\Delta E_1^{\rm 2pt}$ or
fitted, in which case the gap is denoted by $\Delta E$. More
explicitly: Case I uses the energy gap extracted from the two-point
function in both terms in the numerator. In Case II, the two energy
gaps in the numerator are fitted as a common fit parameter. Case III
allows the energy gap $\Delta E$ in the transition term $r_{01}F(\Delta E)$ to vary but
fixes the energy gap of the diagonal term to $\Delta E_1^{\rm
  2pt}$. Case IV truncates the spectral decomposition in the numerator
to include only the transition term $r_{01}F(\Delta E)$ in addition to $g$, thus suppressing the diagonal
term $r_{11}e^{-\Delta E_1^{\rm 2pt}t_s}$, with $\Delta E$ left as a free parameter. Case V further
truncates the spectral decomposition of the denominator of Case IV to
include only the ground state, which allows directly comparing to the
Laplace-filter fit that will be described below.

A potential amplification of the transition amplitude ($r_{01}$)
compared to the diagonal amplitude ($r_{11}$) is accounted for by
fitting using Case~III.  Namely, in this case, the two-point function
energy gap $\Delta E_1^{\rm 2pt}$ is used for the diagonal term in the
numerator, while allowing a potentially current-enhanced transition
term $r_{01}F(\Delta E)$ to have a different energy gap $\Delta E$.

To construct the GEVP-improved ratio $R_{\rm GEVP}^{d}$, we use the weight $W$ in \cref{eq:W} and the GEVP-improved matrix entries
\begin{align}
I_{jk}(t_s,t_{\rm ins})=v_{0j}^*v_{0k}C^{\rm 3pt}_{jk}(t_s,t_{\rm ins}), \label{eq:Ijk}
\end{align}
having suppressed the time arguments for the eigenvectors. We form
\begin{align}
I_{\rm GEVP}^{d}=(1-W^2)I_{NN}+(1+W)\left(I_{N,N\sigma}+I_{N\sigma,N}\right), \label{eq:Igevp}
\end{align}
which is the construction also used in Ref.~\cite{Alexandrou:2024tin}
for the $\pi N$ case. As shown in Ref.~\cite{Alexandrou:2024tin}, in
the two-state limit, this construction cancels the transition contributions proportional to $\braket{0|\mathcal O|1}$ and $\braket{1|\mathcal O|0}$ between
the nucleon and the state isolated by the $N\sigma$ interpolator
without requiring the inclusion of $C^{\rm
  3pt}_{N\sigma,N\sigma}$. The remaining term has an excited state at
both source and sink and is more strongly exponentially suppressed at
large time separations. This construction does not assume that $C^{\rm
  3pt}_{N\sigma,N\sigma}$ itself is small. We form
\begin{align}
R_{\rm GEVP}^{d}(t_s,t_{\rm ins})=\frac{I_{\rm GEVP}^{d}(t_s,t_{\rm ins})}{C_{\rm GEVP}^{\rm 2pt}(t_s)}, \label{eq:Rgevp}
\end{align}
which has the same asymptotic limit as $R_{\rm std}$. When the full matrix of three-point functions is available, the full GEVP-improved construction is
\begin{align}
R_{\rm GEVP}(t_s,t_{\rm ins})=
\frac{\sum_{j,k\in\{N,N\sigma\}}I_{jk}(t_s,t_{\rm ins})}
{C_{\rm GEVP}^{\rm 2pt}(t_s)}. \label{eq:Rgevp-full}
\end{align}
The full GEVP-improved ratio is computationally more demanding for the
case of the light quark insertion, because it requires the connected
contributions to $C^{\rm 3pt}_{N\sigma,N\sigma}$. For the strange and
charm currents, this entry contains no quark connected
insertion, and has been computed as described in
\cref{sec:simulation-details}. We compare the full ratio $R_{\rm
  GEVP}$, using all four matrix entries, and the GEVP-improved ratio
$R_{\rm GEVP}^{d}$ for the strange and charm currents in \cref{sec:strange-charm}.

\begin{figure}[!htbp]
\centering
\includegraphics[width=\columnwidth]{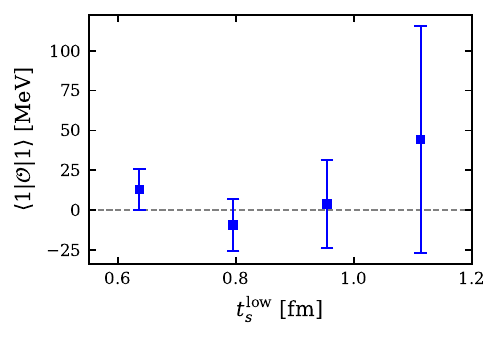}
\caption{Excited-state diagonal matrix element $\braket{1|\mathcal
    O|1}=r_{11}/r_1$ from fits using Case III, versus the minimum
  source--sink separation $t_s^{\rm low}$. The gray dashed line marks
  zero.}
\label{fig:caseIII-matrix-element}
\end{figure}

We compare the analysis of the standard ratio using the fits in the
five cases of \cref{eq:Rstd-two-state-cases} and the GEVP-improved
ratio in \cref{fig:Rstd-Rgevp}. Fitting $R_{\rm std}$ using Case I,
which fixes the excited-state energy gap to the value obtained from
the two-point function fit, yields a lower $\sigma_{\pi N}$ than the
other four fit cases. Cases II and III, which both allow for the
energy gap $\Delta E$ in the transition term $r_{01}F(\Delta E)$ to be fitted rather than fixed by
the two-point function fit, are nearly identical. With Case IV, in
which we omit the diagonal numerator term, the result is similar to
Cases II and III, but with a smaller uncertainty. With Case V, in
which the denominator is truncated to include the ground-state
contribution only, the central values for $\sigma_{\pi N}$ shift
upwards, but are compatible, within the uncertainties, with Cases II
to IV. It is worth noting that with the Case-III fits shown in
\cref{fig:caseIII-matrix-element}, we do not find a value of $r_{11}$
that is non-zero within statistical errors. This suggests that the dominant contamination is
described by the transition term $r_{01}F(\Delta E)$, whose fitted $\Delta E$ can
represent transitions $\braket{0|\mathcal O|n}$ to several excited states $n>0$. The remaining contamination
need not be dominated by $\braket{1|\mathcal O|1}$, but may instead
reflect the inability of a single fitted $\Delta E$ to describe all
these transitions.
Given this, we conclude that Case~IV is the fit that describes our
data best, with $(t_s^{\rm low}/a,t_{\rm ins}^{\rm cut}/a)=(12,2)$,
corresponding to about $t_s\simeq1$~fm. The energy gap obtained is
$\Delta E_1^{\rm 3pt}=408(35)$~MeV, and we will use this value as the
excited state gap in the discussion that follows.

Comparing the standard ratio to the GEVP-improved ratio $R_{\rm
  GEVP}^{d}$ in \cref{fig:Rstd-Rgevp}, we see that the GEVP-improved
ratio has weaker dependence in both $t_{\rm ins}$ and $t_s$ without
loss of precision. This is consistent with a similar analysis at
heavier pion masses in Refs.~\cite{Barca:2024hrl,Barca:2026bmo}. The
midpoint values of $R_{\rm GEVP}^{d}$ shown in the second panel of
\cref{fig:Rstd-Rgevp} agree with the results in the third panel,
i.e. the results obtained after fitting $R_{\rm std}$, for Cases II to
IV, which are the cases in which the energy gaps in the three-point
function are a free fit parameter. On closer inspection though, the
remaining dependence of $R_{\rm GEVP}^{d}$ on $t_{\rm ins}$ and $t_s$
suggests contributions from more than one remaining excited
state. Namely, the two terms that include the transition from
ground to first excited state can be written as
\begin{equation*}
r_{01}F=2r_{01}e^{-\Delta E t_s/2}\cosh\!\left[\Delta E(t_{\rm ins}-t_s/2)\right].
\end{equation*}
For $r_{01}>0$, this form requires $R_{\rm GEVP}^{d}$ for a given
$t_s$ to decrease as $t_{\rm ins}$ approaches the midpoint, $t_s/2$,
which is observed in \cref{fig:Rstd-Rgevp}, but it also requires the
midpoint value to decrease with $t_s$, which is not the case for the
larger values of $t_s$.  This suggests competing residual excited
state contributions, which as we will see is further supported by the
Laplace analysis that will be presented in
\cref{sec:three-point-laplace}. We therefore refrain from fitting
$R_{\rm GEVP}^{d}$ in \cref{fig:Rstd-Rgevp} with either a constant fit
or two-state fits.

\begin{figure}[!htbp]
\centering
\includegraphics[width=\columnwidth]{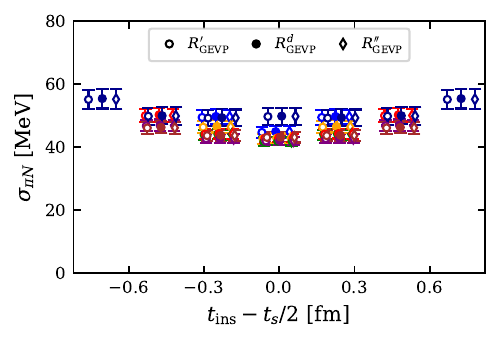}
\caption{GEVP-improved ratio $R_{\rm GEVP}^{d}$ and the variants $R_{\rm GEVP}^{\prime}$ and $R_{\rm GEVP}^{\prime\prime}$, defined in \cref{eq:Rgevp,eq:Rgevp-prime,eq:Rgevp-double-prime}. Colors correspond to different   source--sink time separations with the coloring scheme following the notation  \cref{fig:Rstd-Rgevp}.}
\label{fig:Rgevp-variants}
\end{figure}

To assess the numerical role of the combination in \cref{eq:Igevp}, we
compare $R_{\rm GEVP}^{d}$ with
\begin{align}
I'&=I_{NN}+I_{N,N\sigma}+I_{N\sigma,N},&
R_{\rm GEVP}^{\prime}&=\frac{I'}{C_{\rm GEVP}^{\rm 2pt}}. \label{eq:Rgevp-prime}
\end{align}
We then remove the diagonal $N\sigma$ contribution from the two-point
function in the denominator,
\begin{align}
C_{\rm GEVP}^{\prime\prime\,{\rm 2pt}}&=C_{\rm GEVP}^{\rm 2pt}-|v_{0,N\sigma}|^2C^{\rm 2pt}_{N\sigma,N\sigma},&
R_{\rm GEVP}^{\prime\prime}&=\frac{I'}{C_{\rm GEVP}^{\prime\prime\,{\rm 2pt}}}. \label{eq:Rgevp-double-prime}
\end{align}
The three variants agree as shown in \cref{fig:Rgevp-variants}, as
expected if the leading diagonal contamination is associated with the
first excited state present in the standard nucleon two-point function
rather than the excited state isolated by the $N\sigma$
interpolator. The small changes from the $W$ reweighting and the
diagonal $N\sigma$ contribution to the GEVP-improved denominator are
consistent with this expectation.

To assess whether simpler contractions can approximate the off-diagonal three-point functions at lower computational cost, we compare the full off-diagonal result with two truncations in Appendix~\ref{app:contraction-approximations}: namely, we omit the quark-connected $B,W,Z$ topologies or retain only the current-enhanced contractions. Both reproduce most of the midpoint signal. Their usefulness depends on the target precision, since the omitted terms produce shifts of a few MeV on A48 and B64 and a small relative shift on A24.

\begin{figure*}[!t]
\centering
\includegraphics[width=\textwidth]{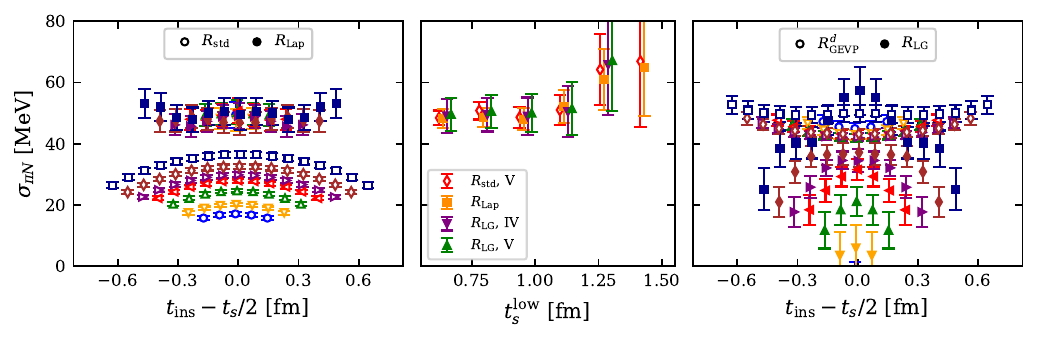}
\caption{Laplace-filter analysis of $\sigma_{\pi N}$. Left: $R_{\rm std}$ and $R_{\rm Lap}$ versus insertion time. Middle: $R_{\rm std}$ case V (open red diamonds), Laplace fits (orange squares), and $R_{\rm LG}$ cases IV (purple downward triangles) and V (green upward triangles) versus $t_s^{\rm low}$. Right: $R_{\rm GEVP}^{d}$ and $R_{\rm LG}$. In the ratio panels, open and filled symbols denote unfiltered and filtered data, with the separation convention of \cref{fig:Rstd-Rgevp}.}
\label{fig:Rstd-Rlap}
\end{figure*}

\subsection{Laplace Filter Method}
\label{sec:three-point-laplace}

The Laplace filter is applied to $t_{\rm ins}$ at fixed $t_s$. This differs from the midpoint application in Ref.~\cite{Portelli:2025lop}, where the filter acts on the source--sink time separation. Using \cref{eq:laplace-filter}, we obtain
\begin{align}
R_{\rm Lap}(t_s,t_{\rm ins};\Delta E)
&=\frac{{\cal L}^{(t_{\rm ins})}_{\Delta E,\delta}C^{\rm 3pt}_{NN}(t_s,t_{\rm ins})}
{{\cal L}^{(t_{\rm ins})}_{\Delta E,\delta}C^{\rm 2pt}_{NN}(t_s)}\nonumber\\
&=\frac{{\cal L}^{(t_{\rm ins})}_{\Delta E,\delta}C^{\rm 3pt}_{NN}(t_s,t_{\rm ins})}
{\lambda^2(\Delta E,\delta)C^{\rm 2pt}_{NN}(t_s)}\nonumber\\
&=\frac{{\cal L}^{(t_{\rm ins})}_{\Delta E,\delta}}{\lambda^2(\Delta E,\delta)}R_{\rm std}(t_s,t_{\rm ins}), \label{eq:Rlap}
\end{align}
where the second equality follows because $C^{\rm 2pt}_{NN}(t_s)$ is
independent of $t_{\rm ins}$. 

We follow the procedure we applied to the two-point function case with
the Laplace filter, namely we fit $R_{\rm Lap}$ to a constant allowing
for $\Delta E$ in \cref{eq:Rlap} to vary. For $(t_s^{\rm low}/a,t_{\rm
  ins}^{\rm cut}/a)=(10,4)$, we obtain
\begin{align}
\Delta E_1^{\rm 3pt,Lap}=360(25)~\text{MeV}. \label{eq:selected-laplace-gap}
\end{align}
Here $\delta/a=2$, meaning $R_{\rm Lap}(t_s,t_{\rm ins};\Delta E)$
depends on $C^{\rm 3pt}_{NN}$ at $t_{\rm ins}-2a$, $t_{\rm ins}$, and
$t_{\rm ins}+2a$. The fit of $R_{\rm Lap}$ over $4a\leq t_{\rm
  ins}\leq t_s-4a$ therefore uses $C^{\rm 3pt}_{NN}$ with $2a\leq
t_{\rm ins}\leq t_s-2a$. In the left panel of \cref{fig:Rstd-Rlap} we
compare the standard ratio with $R_{\rm Lap}$ using $\Delta E_1^{\rm
  3pt,Lap}=360(25)~\text{MeV}$. In the central panel, we compare the
result of fitting $R_{\rm std}$ using Case~V to $R_{\rm Lap}$ as
$t_s^{\rm low}$ is varied.

In the right panel of \cref{fig:Rstd-Rlap} we show the GEVP-improved
ratio ($R_{\rm GEVP}^d$) and compare it to the GEVP-improved ratio
obtained after applying a Laplace filter, namely
\begin{align}
R_{\rm LG}(t_s,t_{\rm ins})=\frac{{\cal L}^{(t_{\rm ins})}_{\Delta E_1^{\rm 3pt,Lap},\delta}I_{\rm GEVP}^{d}(t_s,t_{\rm ins})}{\lambda^2(\Delta E_1^{\rm 3pt,Lap},\delta)C_{\rm GEVP}^{\rm 2pt}(t_s)}. \label{eq:RLG}
\end{align}
Note that we have used $\Delta E_1^{\rm 3pt,Lap}=360(25)~\text{MeV}$
determined from the Laplace method fit to the standard ratio, since
fitting the Laplace filtered $R_{\rm GEVP}^d$ leaving the energy gap
to vary leads to an unstable fit, suggesting that the time dependence
of $R_{\rm GEVP}^d$ may require more than one excited state to be
robustly described. Indeed, as can be seen in the right panel of
\cref{fig:Rstd-Rlap}, filtering using $\Delta E_1^{\rm 3pt,Lap}$
yields $R_{\rm LG}$ with a time dependence more pronounced than the
ratio we started from.

Fitting to $R_{\rm LG}$ using Cases IV and V yields results for the
sigma-term compatible with the fits to $R_{\rm std}$ and to $R_{\rm
  Lap}$, as shown in the central panel of \cref{fig:Rstd-Rlap}. We
retain Case IV, using the standard two-point function in the
denominator, at $(t_s^{\rm low}/a,t_{\rm ins}^{\rm
  cut}/a)=(10,4)$. The energy gap obtained in this case is $\Delta
E_1^{\rm LG}=850(190)$~MeV.

\begin{figure*}[!t]
\centering
\includegraphics[width=\textwidth]{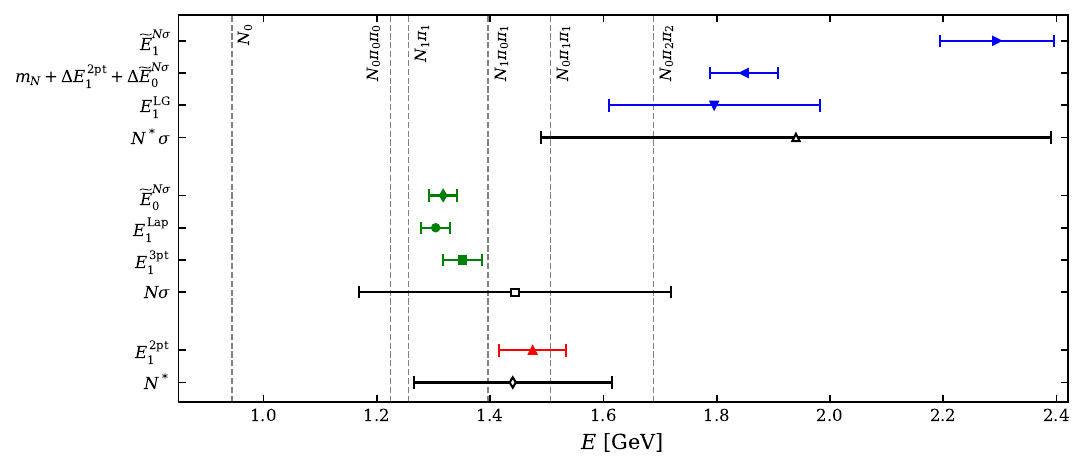}
\caption{Comparison of energy scales from the two- and three-point analyses. Red denotes $E_1^{\rm 2pt}$, green denotes $E_1^{\rm 3pt}$, $E_1^{\rm Lap}$, and $\widetilde E_0^{N\sigma}$, and blue denotes $\widetilde E_1^{N\sigma}$, $m_N+\Delta E_1^{\rm 2pt}+\Delta\widetilde E_0^{N\sigma}$, and $E_1^{\rm LG}$. Vertical gray dashed lines show noninteracting levels evaluated with the ensemble masses and volume. Momentum subscripts $k$ denote $|\boldsymbol p|^2=k\,(2\pi/L)^2$. The open black points are PDG-based $N^*$, $N\sigma$, and $N^*\sigma$ reference energies~\cite{ParticleDataGroup:2024cfk}, with $N^*\equiv N^*(1440)$ and $\sigma\equiv f_0(500)$. The $N\sigma$ reference uses our nucleon mass. Their horizontal bars represent half-width ranges.}
\label{fig:energy-scales}
\end{figure*}

The different behavior of the standard and GEVP-improved ratios can be
understood if we analyze the effect of the Laplace filter on a ratio
with more than one excited state. In general, for the transition term
between the ground state and an excited state $n$, with energy gap
$\Delta E_n=E_n-m_N$, the Laplace filter introduces a term that
multiplies the transition matrix element $\braket{0|\mathcal O|n}$ by
\begin{align}
f_{\rm Lap}(\Delta E_n;\Delta E,\delta)=1-\frac{\lambda^2(\Delta E_n,\delta)}{\lambda^2(\Delta E,\delta)}. \label{eq:laplace-transition-response}
\end{align}
Using $\Delta E_1^{\rm LG}=850$~MeV for $\Delta E_n$ and the value
$\Delta E_1^{\rm 3pt,Lap}=360$~MeV, for $\Delta E$ in
\cref{eq:laplace-transition-response} gives $f_{\rm Lap}(\Delta
E_1^{\rm LG};\Delta E_1^{\rm 3pt,Lap},2a)\simeq-4.8$. Applying the
Laplace filter to a ratio with three states thus can potentially
reverse the time dependence of the ratio, which is what we observe in
\cref{fig:Rstd-Rlap}, namely that at fixed $t_s$, $R^d_{\rm GEVP}$
decreases toward $t_{\rm ins}=t_s/2$, whereas after the Laplace filter
is applied to yield $R_{\rm LG}$, this time dependence is flipped and
amplified. This picture is compatible with a relatively flat
GEVP-improved ratio being the result of opposing residual
contributions and partial cancellation rather than negligible
excited-state contamination, which is revealed after a Laplace filter
eliminates the contribution with energy gap $\Delta E_1^{\rm 3pt,Lap}$ as in
\cref{eq:laplace-transition-response}.

We can further investigate this three-state model using a schematic in
which we have two interpolating operators, namely ${\cal J}_N$ and
$\widetilde{\cal J}_{N\sigma}$, and assume that both couple to a
common excited state $|X\rangle$, with $\widetilde{\cal J}_{N\sigma}$
also coupling to a higher energy contribution $|Y\rangle$. With
$|0\rangle$ denoting the nucleon ground state and assuming
$\widetilde{\cal J}_{N\sigma}$ is orthogonal to $|0\rangle$ we
normalize each interpolator by its leading overlap,
\begin{align}
{\cal J}_N^\dagger|\Omega\rangle
&\propto |0\rangle+\sqrt{r_1^N}|X\rangle, \nonumber\\
\widetilde{\cal J}_{N\sigma}^\dagger|\Omega\rangle
&\propto |X\rangle+\sqrt{r_1^{\widetilde{N\sigma}}}|Y\rangle, \label{eq:schematic-overlaps}
\end{align}
where $r_1^N=|Z_{N,1}/Z_{N,0}|^2$ and
$r_1^{\widetilde{N\sigma}}=|Z_{\widetilde{N\sigma},1}/Z_{\widetilde{N\sigma},0}|^2$.
The optimal nucleon interpolating operator ($\widetilde{\cal J}_N$) in
this schematic would be the one that eliminates the contribution from
$| X\rangle$, which however would leave a residual contribution
from $|Y\rangle$:
\begin{align}
\widetilde{\cal J}_N^\dagger|\Omega\rangle
&\propto |0\rangle-\sqrt{r_1^N r_1^{\widetilde{N\sigma}}}|Y\rangle. \label{eq:schematic-gevp-improved-operator}
\end{align}
Using the values of $r_1^N$ and $r_1^{\widetilde{N\sigma}}$ from the
GEVP analysis of the two-point function in
\cref{sec:two-point-analysis}, the residual squared overlap is
\begin{align}
r_1^N r_1^{\widetilde{N\sigma}}=2.24(21). \label{eq:gevp-inherited-overlap}
\end{align}
Thus, cancelling the common contribution $|X\rangle$ can leave a sizeable overlap with $|Y\rangle$ through $\widetilde{\cal J}_{N\sigma}$. In this model, $|Y\rangle$ represents the first excited contribution in the GEVP-improved $N\sigma$ two-point function, with fitted energy $\widetilde E_1^{N\sigma}$.

\begin{figure}[!htbp]
\centering
\includegraphics[width=\columnwidth]{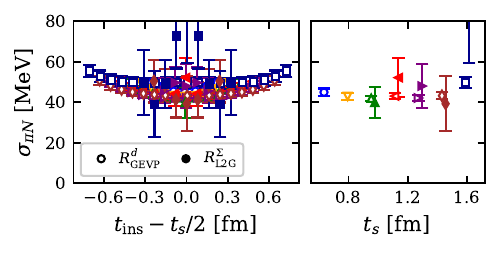}
\caption{GEVP-improved (open) and double-filtered (filled) light-current ratios, versus insertion time (left) and at the midpoint versus $t_s$ (right). Colors follow \cref{fig:Rstd-Rgevp}.}
\label{fig:RL2G-energy-comparison}
\end{figure}

We compare the excited state energies obtained from fits to $R_{\rm
  std}$ and $R_{\rm Lap}$ with those obtained from fits to the
two-point functions. We define $E_1^{\rm 3pt}=m_N+\Delta E_1^{\rm
  3pt}$ as the excited state energy obtained from fits to the standard
ratio, and $E_1^{\rm Lap}=m_N+\Delta E_1^{\rm 3pt,Lap}$, as the
excited state energy obtained from the Laplace filter applied to the
standard ratio. As can be seen in \cref{fig:energy-scales}, both
energies lie close to the energy obtained from the GEVP analysis of
the two-point function, $\widetilde E_0^{N\sigma}$, and below the
first excited state from the fit to the standard nucleon two-point
function, $E_1^{\rm 2pt}$. In \cref{fig:energy-scales}, we also show
energy levels of noninteracting positive-parity nucleon-pion states,
with momenta denoted via subscript, i.e.  the subscript $k$ denotes
that state with momentum $|\vec p|^2=k\,(\frac{2\pi}{L})^2$.

Fitting the filtered GEVP-improved ratio $R_{\rm LG}$, the energy is
$E_1^{\rm LG}=m_N+\Delta E_1^{\rm LG}=1.79(19)$~GeV, which is
compatible with $m_N+\Delta E_1^{\rm 2pt}+\Delta\widetilde
E_0^{N\sigma}$, but lies below $\widetilde
E_1^{N\sigma}=2.29(10)$~GeV. The latter is the energy assigned to
$|Y\rangle$ in the schematic model, so this assignment does not
directly explain the energy extracted from $R_{\rm LG}$.

Indeed, the schematic in \cref{eq:gevp-inherited-overlap} illustrates how a
higher-state contribution can remain after GEVP improvement, but when
several contributions present in the correlation functions are
represented by a single excited state in the fit, it may not be
possible to interpret the energy extracted from $R_{\rm LG}$.

The open black points in \cref{fig:energy-scales} show reference energies constructed from the PDG values for the Roper resonance $N^*\equiv N^*(1440)$ and $\sigma\equiv f_0(500)$: $(M_{N^*},\Gamma_{N^*})=(1.44,0.35)$~GeV and $(M_\sigma,\Gamma_\sigma)=(0.50,0.55)$~GeV~\cite{ParticleDataGroup:2024cfk}. The $N\sigma$ reference combines $M_\sigma$ with our nucleon mass.

This picture motivates applying a Laplace filter to $R_{\rm LG}$ with
energy gap $\Delta E_\Sigma=\Delta E_1^{\rm 2pt}+\Delta E_1^{\rm
  3pt,Lap}$. Starting from $R^d_{\rm GEVP}$, the ratio is given by
\begin{align}
R_{\rm L2G}^{\Sigma}=\frac{{\cal L}_{\Delta E_\Sigma,\delta_2}{\cal L}_{\Delta E_1^{\rm 3pt,Lap},\delta_1}I_{\rm GEVP}^{d}}{\lambda^2(\Delta E_\Sigma,\delta_2)\lambda^2(\Delta E_1^{\rm 3pt,Lap},\delta_1)C_{\rm GEVP}^{\rm 2pt}}. \label{eq:RL2G}
\end{align}
We compare the results using this double-filtered ratio with those
using $R_{\rm GEVP}^{d}$ in \cref{fig:RL2G-energy-comparison}. As can
be seen, the result is too noisy to draw a definite conclusion about
the remaining excited-state contamination.

To summarize, for the light sigma-term, GEVP improvement substantially
reduces the visible excited-state contamination while maintaining the
statistical precision of the ratio. If the residual contamination were
negligible, this would allow a more precise and reliable extraction of
$\sigma_{\pi N}$. Here, however, residual contributions and the
higher-state contamination exposed by filtering prevent such an
extraction from the nearly flat ratio. Accounting for these effects by
fitting the filtered ratio $R_{\rm LG}$ using cases IV and V gives
larger uncertainties than the results obtained when using the ratio
with the standard nucleon interpolator. Thus, the $2\times2$ GEVP
considered in this work provides a useful cross-check but does not
improve the extraction of $\sigma_{\pi N}$.

\subsection{Comparison of analysis methods}
\label{sec:light-comparison}

\begin{figure}[tb]
\centering
\includegraphics[width=\columnwidth]{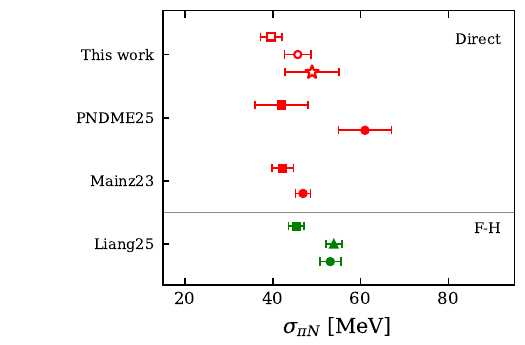}
\caption{Dependence of $\sigma_{\pi N}$ on the analysis method. Red and green denote direct and F-H-based analyses. For this work, squares and circles show cases I and IV for the ratio without GEVP, $R_{\rm std}$ of \cref{eq:Rstd}; stars show case IV for the Laplace-filtered GEVP-improved ratio $R_{\rm LG}$ of \cref{eq:RLG}. For PNDME25, squares and circles denote wide-prior and narrow $N\pi$/$N\pi\pi$-prior fits~\cite{Park:2025rxi}; for Mainz23, summation and two-state $N\pi\pi$-prior fits~\cite{Agadjanov:2023efe}. For Liang25, squares, triangles, and circles denote one-loop $\mathcal O(p^3)$, one-loop $\mathcal O(p^4)$, and two-loop $\mathcal O(p^5)$ fits~\cite{Liang:2025adz}. Open and filled lattice symbols denote finite-spacing and continuum-extrapolated results.}
\label{fig:sigma-piN-methods}
\end{figure}

In~\cref{fig:sigma-piN-methods}, we compare results for $\sigma_{\pi N}$ obtained using different analysis methods within this work and in Refs.~\cite{Park:2025rxi,Agadjanov:2023efe,Liang:2025adz}.

In the F-H approach, $\sigma_{\pi N}$ is directly related to the slope of the nucleon mass with respect to the light-quark mass, making the dependence of the nucleon on the pion mass an important aspect. For example, the results denoted by ``Liang25'' in \cref{fig:sigma-piN-methods} are from one- and two-loop chiral analyses of the Mainz lattice results~\cite{Liang:2025adz}. At the neutral pion mass, the one-loop $\mathcal O(p^3)$ and $\mathcal O(p^4)$ fits give $45.4(1.8)$ and $53.9(1.8)$~MeV, while the two-loop $\mathcal O(p^5)$ result is $53.1(2.4)$~MeV. The close $\mathcal O(p^4)$ and $\mathcal O(p^5)$ central values do not imply negligible two-loop effects, since the $\mathcal O(p^4)$ expression includes fit parameters that can absorb part of these effects. The two-loop calculation includes isoscalar $\pi\pi$ rescattering, related to matrix elements with $N\pi\pi$ states, which may contribute to the scalar three-point function calculation.

Low-lying $N\pi$ and $N\pi\pi$ contributions can be enhanced in the scalar three-point function while remaining difficult to resolve in the nucleon two-point function, making the extracted value sensitive to contamination from excited-states. The two results labeled ``PNDME25'' in \cref{fig:sigma-piN-methods} illustrate this dependence. Namely, when fitting with a narrow prior centered at the lowest noninteracting $N\pi$ or $N\pi\pi$ energy, they find $61(6)$~MeV for the sigma-term (circle), while using wide priors in their standard analysis, they find $42(6)$~MeV (square)~\cite{Park:2025rxi}. The two nearly degenerate contributions are represented by one effective excited state.

In the analysis of Ref.~\cite{Agadjanov:2023efe}, labeled ``Mainz23'' in \cref{fig:sigma-piN-methods}, a smaller effect is identified between their result of $42.3(2.4)$~MeV when using the summation method, and their value of $46.9(1.7)$~MeV when a two-state fit is performed with a prior centered at the noninteracting $N\pi\pi$ energy gap of $2m_\pi$. The increase of about $5$~MeV is similar in size to the shift observed in our fits to $R_{\rm std}$ when comparing case I, which fixes both numerator energy gaps to $\Delta E_1^{\rm 2pt}$ from the nucleon two-point correlator, and case IV, in which we fit $\Delta E_1^{\rm 3pt}$ and omit the unresolved diagonal term. These results expose similar fit-model dependences, although the fit strategies differ.

The effectiveness of a variational analysis depends on the
interpolators included. Our earlier GEVP using $N\pi$ interpolating
operators gave little improvement for the scalar
current~\cite{Alexandrou:2024tin}. This is also the case for the results of 
Ref.~\cite{Wang:2025nsd}, where the isovector scalar and tensor ratios had similar time dependence with and without an $N$--$N\pi$ GEVP at moderate and large source--sink separations. Instead, they combined the nucleon interpolator with an additional interpolator formed by multiplying it by the spatially summed scalar or tensor current to reduce excited-state effects.

For the  isoscalar scalar light quark operator, an $N-N\sigma$ basis targets a
different channel. In the off-diagonal $N$--$N\sigma$ three-point
function, the current can couple to the scalar-isoscalar part of the
$N\sigma$ interpolator, while the nucleon propagates separately. This
current--meson contraction is quark-line disconnected from the nucleon
and can be enhanced by the spatial volume at matching momenta~\cite{Barca:2025det}.
This current enhancement can make the ground-to-excited-state transition $\braket{0|\mathcal O|n}$ prominent in the three-point
function even if the corresponding state has little weight in the
nucleon two-point function. Near the physical pion mass, the
scalar-isoscalar channel also couples to $S$-wave $\pi\pi$ states.
Using an $N\sigma$ basis, 
excited-state contamination at heavier pion masses was suppressed~\cite{Barca:2024hrl,Barca:2026bmo}.
In contrast in our calculation at physical pion mass, although the GEVP-improved  ratio for $\sigma_{\pi N}$ is flatter,
the residual contributions discussed above prevent us from treating
that flatness alone as ground-state dominance. For our analyses shown in \cref{fig:sigma-piN-methods}, we use $t_s^{\rm low}\approx 1$~fm. The case-IV fit to the filtered ratio $R_{\rm LG}$ gives $49.0(6.1)$~MeV. This raises the central value relative to the case-IV fit to $R_{\rm std}$, but with a substantially larger statistical error. The GEVP result therefore serves as a cross-check rather than a better extraction of $\sigma_{\pi N}$.

For our final result, we take the central value and statistical error from the case-IV fit to the standard ratio $R_{\rm std}$. The systematic uncertainty due to the excited-state analysis is computed by taking the absolute difference between this central value and that obtained from the case-IV fit to the filtered ratio $R_{\rm LG}$. This gives
\begin{equation}
\sigma_{\pi N}=45.7(3.0)(3.3)~\mathrm{MeV}.
\label{eq:sigma-piN-result}
\end{equation}

\begin{figure*}[!t]
\centering
\includegraphics[width=\textwidth]{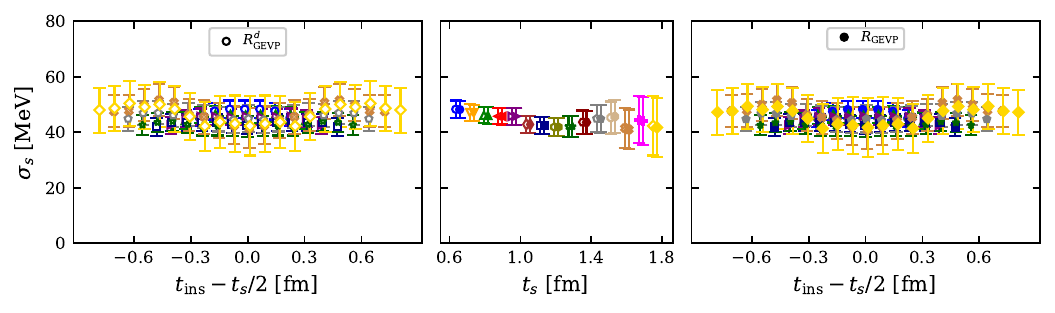}
\caption{Comparison of $R_{\rm GEVP}^{d}$ and the full GEVP-improved ratio $R_{\rm GEVP}$ for the strange current. Left and right: insertion-time dependence of $R_{\rm GEVP}^{d}$ (open) and $R_{\rm GEVP}$ (filled), respectively, at even separations. Middle: midpoint values including odd and even separations, averaging the two central insertions for odd $t_s/a$. Colors and symbols identify the same $t_s$ throughout.}
\label{fig:sigma-s-full}
\end{figure*}

\begin{figure*}[!t]
\centering
\includegraphics[width=\textwidth]{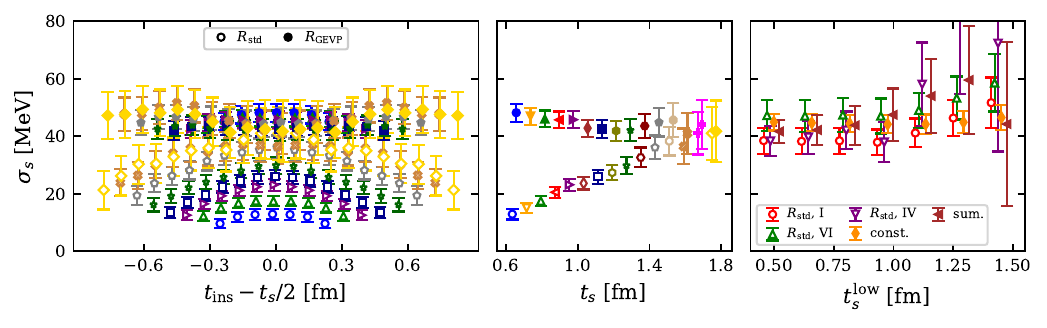}
\caption{Results on the ratio for the extraction of $\sigma_s$. The first two panels compare the insertion-time dependence and midpoint values of $R_{\rm std}$ (open symbols) and $R_{\rm GEVP}$ (filled symbols), using the color convention of \cref{fig:sigma-s-full}. The third panel shows results versus $t_s^{\rm low}$ using the ratio with the standard nucleon interpolator for cases I (red), IV (purple), and VI (green), and GEVP-improved ratio using constant (orange) and summation (brown) fits.}
\label{fig:sigma-s-gevp}
\end{figure*}

\section{Strange- and Charm-Quark Sigma Terms}
\label{sec:strange-charm}

The analyses for the strange and charm sigma-terms are  extensions of our analysis for $\sigma_{\pi N}$. Data are available at every integer separation $2\leq t_s/a\leq30$, including odd separations and larger values of  $t_s$. We also have the $C^{\rm 3pt}_{N\sigma,N\sigma}$ entry, allowing a direct comparison of the GEVP-improved ratios $R_{\rm GEVP}^{d}$ and $R_{\rm GEVP}$. The scalar operators use the isoQCD bare quark masses specified in \cref{sec:pion-mass-corrections}.

\begin{figure}[h!]
\centering
\includegraphics[width=\columnwidth]{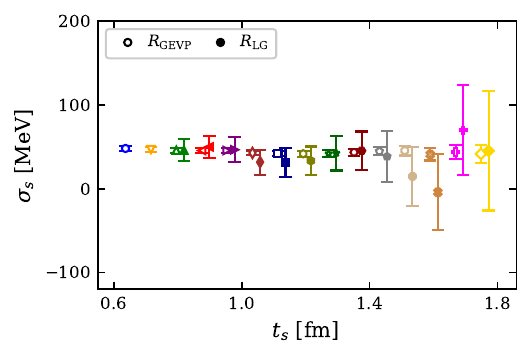}
\caption{Midpoint values of the full GEVP-improved strange-current ratio
  $R_{\rm GEVP}$ (open) and its Laplace-filtered counterpart $R_{\rm LG}$ (filled)
  versus $t_s$. The filter uses the gap
  $\Delta E_1^{\rm 3pt,Lap}$ determined from the $\sigma_{\pi N}$ analysis.
  Colors and symbols follow
  \cref{fig:sigma-s-full}.}
\label{fig:sigma-s-laplace}
\end{figure}

For clarity, we do not show all the analyzed data in the insertion-time
and midpoint plots. The midpoint is the value of the ratio at $t_{\rm
  ins}=t_s/2$, for even $t_s/a$ and the average between $t_{\rm
  ins}=(t_s-a)/2$ and $t_{\rm ins}=(t_s+a)/2$ for odd $t_s/a$.
\subsection{Strange sigma-term}
\label{sec:strange-sigma}

For the strange sigma-term $\sigma_s$, we compare the GEVP-improved ratio $R_{\rm
  GEVP}^{d}$ of \cref{eq:Rgevp} with the full ratio $R_{\rm GEVP}$ of
\cref{eq:Rgevp-full} in \cref{fig:sigma-s-full}. As can be seen, we
observe agreement with comparable statistical precision. We use the full ratio $R_{\rm GEVP}$ for the extraction of
$\sigma_s$.

Unlike $R_{\rm std}$, $R_{\rm GEVP}$ shows generally no dependency on
both insertion time and source--sink time separation, as shown in
\cref{fig:sigma-s-gevp}. The effectiveness of the GEVP with the
$N-N\sigma$ basis in identifying the ground state was observed also at
a heavier pion mass in Ref.~\cite{Barca:2024hrl}. To compare the
values extracted with and without GEVP, we show two-state fits to
$R_{\rm std}$ alongside constant fits and fits using the summation
method for $R_{\rm GEVP}$. In the summation method, we form $S_s$,
\begin{align}
\mathcal S_s(t_s)=a\sum_{n=1}^{t_s/a-1}R_{\rm GEVP}(t_s,na), \label{eq:strange-summation}
\end{align}
and fit to $\mathcal S_s(t_s)=c+\sigma_s t_s$ using the same $t_s$ as
for the constant fits.  For the ratio using the standard nucleon
interpolator, we use cases I and IV of \cref{eq:Rstd-two-state-cases}
and a further case, which we will refer to as case VI, in which the
energy gap $\Delta E$ is set to $\Delta E_1^{\rm 3pt}$ in case III,
fixed to the value obtained from the case-IV fit to $R_{\rm std}$ in
the analysis of $\sigma_{\pi N}$. The resulting ratio is given by
\begin{equation}
R_{\rm VI}=\frac{g+r_{01}F(\Delta E_1^{\rm 3pt})+r_{11}e^{-\Delta E_1^{\rm 2pt}t_s}}
{1+r_1e^{-\Delta E_1^{\rm 2pt}t_s}}. \label{eq:Rstd-two-state-VI}
\end{equation}
Its parameters $g$, $r_{01}$, and $r_{11}$ are fitted, with the
two-point inputs treated as in case III. The value extracted using
$R_{\rm VI}$ becomes less precise as $t_s^{\rm low}$ increases,
whereas the constant fits to $R_{\rm GEVP}$ remain stable. The
summation results agree with the value extracted with the constant
fits, providing a cross-check albeit carrying larger uncertainties.

We also tested cases II and III with $\Delta E$ as a free fit
parameter and the Laplace method. Cases II and III are poorly
constrained and vary strongly with the fit range used. In
\cref{fig:sigma-s-laplace} we compare $R_{\rm GEVP}$ with its
Laplace-filtered counterpart $R_{\rm LG}$, using as the filter energy $\Delta
E_1^{\rm 3pt,Lap}$ determined from the analysis of $\sigma_{\pi
  N}$. The filter amplifies statistical fluctuations, leaving the
strange ratio too noisy to distinguish residual excited-state
contributions.

The nearly flat behavior of $R_{\rm GEVP}$ over the time separations,
together with the stability of the fits when using different fit
ranges in $t_s^{\rm low}$, suggests one-state dominance and thus a
constant fit for extracting $\sigma_s$. For $t_s^{\rm
  low}\approx0.8$~fm, the constant fit gives $44.4(3.0)$~MeV.

For our final result, we take the central value and statistical error from the case-IV fit to the standard ratio $R_{\rm std}$ at $t_s^{\rm low}\approx0.8$~fm in \cref{fig:sigma-s-gevp}. The systematic uncertainty due to the excited-state analysis is computed by taking the absolute difference between this central value and that obtained from the constant fit to $R_{\rm GEVP}$. This gives
\begin{equation}
\sigma_s=42.1(6.5)(2.3)~\mathrm{MeV}.
\label{eq:sigma-s-result}
\end{equation}

\begin{figure*}[!t]
\centering
\includegraphics[width=\textwidth]{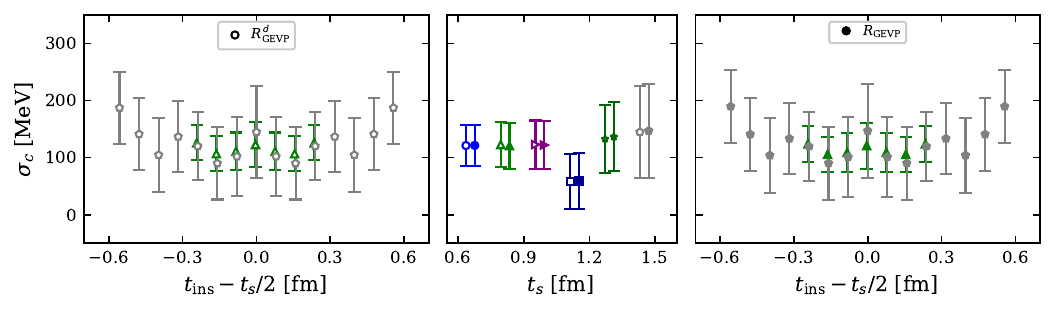}
\caption{Comparison of $R_{\rm GEVP}^{d}$ and the full GEVP-improved ratio $R_{\rm GEVP}$ for the charm current. Left and right: insertion-time dependence of $R_{\rm GEVP}^{d}$ (open) and $R_{\rm GEVP}$ (filled), respectively. Middle: midpoint values. Colors and symbols follow \cref{fig:sigma-s-full}.}
\label{fig:sigma-c-full}
\end{figure*}

\subsection{Charm sigma-term}
\label{sec:charm-sigma}

For the charm sigma-term $\sigma_c$, we compare the GEVP-improved
ratios $R_{\rm GEVP}^{d}$ and $R_{\rm GEVP}$ in
\cref{fig:sigma-c-full}. They agree with comparable statistical
precision. We use the full ratio $R_{\rm GEVP}$ for the following
analysis.

\begin{figure*}[!t]
\centering
\includegraphics[width=\textwidth]{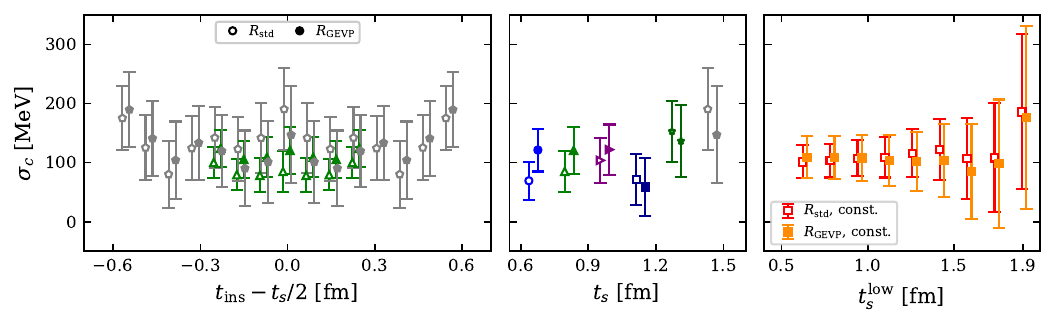}
\caption{Results on the ratios for the extraction of $\sigma_c$. The first two panels compare the insertion-time dependence and midpoint values of $R_{\rm std}$ (open symbols) and $R_{\rm GEVP}$ (filled symbols). The third panel shows constant-fit results versus $t_s^{\rm low}$ for $R_{\rm std}$ (open red squares) and $R_{\rm GEVP}$ (filled orange squares). For clarity, only selected separations are displayed in the first two panels, while the fits use all available data within the corresponding fit ranges.}
\label{fig:sigma-c-gevp}
\end{figure*}

In \cref{fig:sigma-c-gevp}, we compare constant fits to $R_{\rm std}$
and $R_{\rm GEVP}$. We use $t_{\rm ins}^{\rm cut}=4a$ for both
ratios. For $t_s^{\rm low}\approx1$~fm, these fits give $107.4(30.4)$
and $107.9(38.5)$~MeV, respectively.

For our result, we take the central value and statistical error from
the constant fit to $R_{\rm std}$ with $t_s^{\rm low}\approx1$~fm in
\cref{fig:sigma-c-gevp}. The systematic uncertainty due to the
excited-state analysis is computed by taking the absolute difference
between this central value and that obtained from the constant fit to
$R_{\rm GEVP}$. This gives
\begin{equation}
\sigma_c=107.4(30.4)(0.6)~\mathrm{MeV}.
\label{eq:sigma-c-result}
\end{equation}

\section{Conclusions}
\label{sec:conclusions}

In this work, we study the effect of excited-states in the extraction of  the light,
strange, and charm nucleon sigma-terms. We consider a basis comprising the standard nucleon and
nucleon--$\sigma$-meson interpolating operators. We use an ensemble of
$N_f=2+1+1$ twisted-mass clover-improved fermions at the physical pion
mass, complemented by an $N_f=2$ ensemble at the physical pion
mass and an $N_f=2+1+1$ ensemble with $m_\pi=362$~MeV. In all cases, we use the nucleon lab frame.   

For the analysis of the two-point functions we use multi-state fits, a
GEVP analysis, and the Laplace filter method. Two-state fits to the 
nucleon effective mass constructed using the standard nucleon interpolator give $m_N=944.0(4.4)$~MeV and an energy gap of $\Delta E_1^{\rm
  2pt}=531(55)$~MeV. Applying the Laplace-filter yields results in agreement.
The GEVP eigenvector ratio $v_{0,N\sigma}/v_{0,N}=0.01564(99)$ leaves
the nucleon effective mass nearly unchanged, while the GEVP-improved
$N\sigma$ correlator yields $\widetilde E_0^{N\sigma}=1.317(25)$~GeV
and $\widetilde E_1^{N\sigma}=2.29(10)$~GeV.

To extract the $\sigma_{\pi N}$ using the B64 ensemble, we constructed a GEVP-improved ratio
that does not require the three-point function
$\langle N\sigma|\bar{u}u+\bar{d}d|N\sigma\rangle$, which is not
available to us. We also apply the Laplace filter with respect to the
insertion time to both the ratio without GEVP improvement,
$R_{\rm std}$ in \cref{eq:Rstd}, and to the ratio with GEVP improvement,
$R_{\rm GEVP}^{d}$ in \cref{eq:Rgevp}.

The GEVP-improved  ratio using the  B64 ensemble exhibits substantially reduced
time dependence. Applying the Laplace filter, however, points to residual excited-state
contributions. Using the A24 and A48 ensembles, the GEVP improvement similarly reduces the
 dependence on the  source--sink time separation. Applying the Laplace filter again reveals residual time separation dependence for the 
A24 ensemble. A similar but milder effect is also observed when using the A48 ensemble. 

For the strange and charm three-point functions, data
for both the standard and GEVP-improved ratios are available for all
source-sink separations $2\leq t_s/a\leq30$.
In these cases, we also have the full $2\times2$ correlator matrix for the GEVP analysis. For the strange sigma-term, the
constant fit to the full GEVP-improved ratio with $t_s^{\rm low}\approx0.8$~fm gives a value of 
$44.4(3.0)$~MeV, consistent with the result obtained using the
summation method. For the charm sigma-term, constant fits to $R_{\rm std}$ or $R_{\rm GEVP}$ give compatible values.

In general, this work shows that the effect of adding the  $\sigma
N$ interpolating operator depends on the flavor of the scalar current
under investigation:
\begin{itemize}
\item $\sigma_{\pi N}$: The GEVP-improved ratio has substantially reduced time dependence without loss of statistical precision. Our results with GEVP and the Laplace filter support partial cancellation of residual excited-state contributions at the expense of increasing statistical noise. The present analysis, therefore, provides a useful cross-check, but does not yield a more precise determination of $\sigma_{\pi N}$ as compared to the two-state analysis. Thus, as our final value for the B64 ensemble, we quote $\sigma_{\pi N}=45.7(3.0)(3.3)$~MeV, as determined from the fit to $R_{\rm std}$ of case-IV. The systematic error is given as the absolute difference between the mean values determined when we fit to $R_{\rm std}$ and to $R_{\rm LG}$ both for case-IV.
\item $\sigma_s$: The full GEVP-improved ratio shows reduced time dependence on the insertion and source--sink separation times, enabling stable constant fits with better statistical precision than fits to the GEVP-unimproved $R_{\rm std}$. Since the values are compatible, for consistency with using the value from the two-state fit  for $\sigma_{\pi N}$,  we quote as our final value for the B64 ensemble $\sigma_s=42.1(6.5)(2.3)$~MeV, determined from fitting to $R_{\rm std}$, where the systematic error is again determined from the absolute difference in the mean values from fitting to $R_{\rm std}$ and to $R_{\rm GEVP}$.
\item $\sigma_c$: The unimproved ratio already shows only mild time dependence, with little change after applying GEVP improvement. Thus, for our final value we  quote $\sigma_c=107.4(30.4)(0.6)$~MeV from the constant fit to $R_{\rm std}$ for the B64 ensemble, with the same procedure for the determination of the systematic error.
\end{itemize}

This motivates further variational studies, in particular with an
enlarged basis to resolve the residual contamination identified for
$\sigma_{\pi N}$.

\begin{acknowledgments}
We would like to thank all members of the Extended Twisted Mass
Collaboration for a very enjoyable cooperation.
C.A., G.K., Y.L., S.P. and
F.P. acknowledge partial support from the projects 3D-nucleon
(EXCELLENCE/0421/0043), IMAGE-N (EXCELLENCE/0524/0459), MuonHVP
(EXCELLENCE/0524/0017), PulseQCD (EXCELLENCE/0524/0269), Baryon8
(EXCELLENCE/0524/0001), RobustSigma (EXCELLENCE/0925/0325),
DeNuTra (EXCELLENCE/524/0455), LENA-QCD (POST-DOC/0925 - ROandOther/0180) and partonWF (VISION ERC/0525/0010)
funded by the European Regional Development Fund and the Republic of
Cyprus through the Cyprus Research and Innovation Foundation. C.A. is
supported by European Union's HORIZON MSCA Doctoral Networks
programme, under Grant Agreement No. 101072344, project AQTIVATE
(Advanced computing, QuanTum algorIthms and data-driVen Approaches for
science, Technology and Engineering).
The authors gratefully acknowledge the Gauss Centre for Supercomputing
e.V. (www.gauss-centre.eu) for funding this project by providing
computing time through the John von Neumann Institute for Computing
(NIC) on the GCS Supercomputers JUWELS, JUWELS
Booster~\cite{JUWELS-BOOSTER} and JUPITER Booster at J\"ulich
Supercomputing Centre (JSC). We also acknowledge computing time
granted on Piz Daint and Alps at Centro Svizzero di Calcolo Scientifico (CSCS)
via the projects s1174 and lp139.
\end{acknowledgments}

\bibliography{references}

\appendix

\section{Light-Current GEVP Analysis on A24 and A48}
\label{app:a24-a48}

\begin{figure*}[!tbp]
\centering
\includegraphics[width=\textwidth]{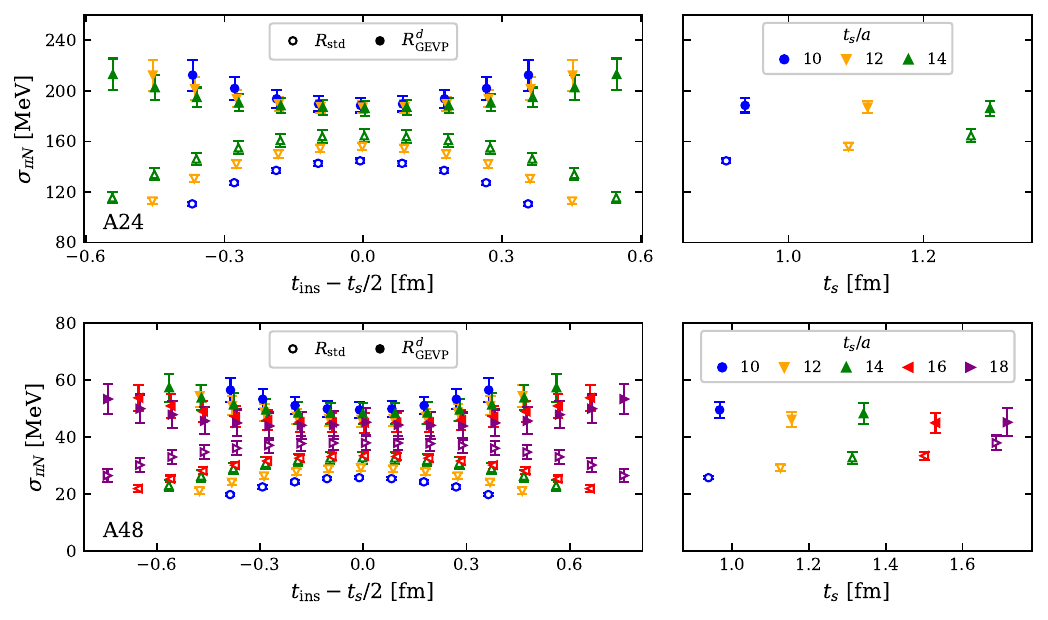}
\caption{Standard (open) and GEVP-improved (filled) light-current ratios on A24 (top) and A48 (bottom). Left: insertion-time dependence. Right: midpoint values versus $t_s$. Colors and symbols identify the source--sink separations within each row.}
\label{fig:a-ensembles-standard-gevp}
\end{figure*}

\begin{figure*}[!tbp]
\centering
\includegraphics[width=\textwidth]{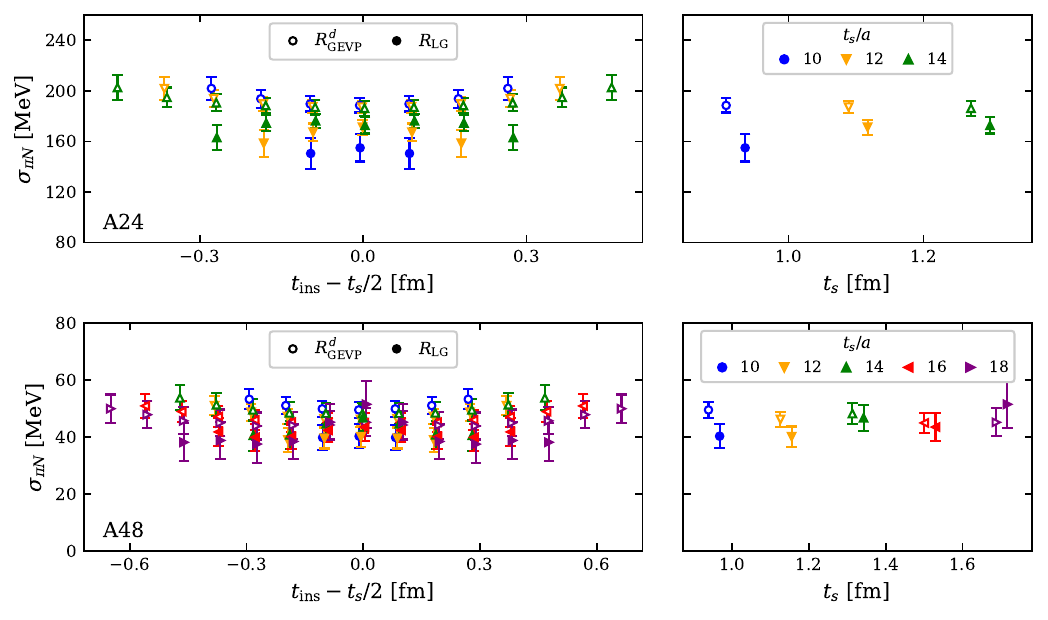}
\caption{GEVP-improved (open) and Laplace-filtered (filled) light-current ratios on A24 and A48, with the panels and separation conventions of \cref{fig:a-ensembles-standard-gevp}. The filter uses the energy gap $\Delta E_1^{\rm 3pt,Lap}$ determined from $R_{\rm Lap}$ on the corresponding ensemble, including its uncertainty.}
\label{fig:a-ensembles-gevp-laplace}
\end{figure*}

We extend the light-current comparison to the ensemble with heavier
than physical pion mass, A24, and the physical-point ensemble A48,
using the variational basis of \cref{sec:two-point-gevp} and the
standard and GEVP-improved ratios defined in
\cref{eq:Rstd,eq:Rgevp}. For the A24 sigma terms, we multiply the bare
scalar matrix elements with the simulated bare light-quark mass, while
for A48 we use the leading isoQCD mass replacement of
\cref{sec:pion-mass-corrections}.

In \cref{fig:a-ensembles-standard-gevp}, we compare the standard and
GEVP-improved ratios. On both ensembles, GEVP improvement shifts the
ratios upward and substantially reduces their source--sink-separation
dependence. This is especially clear in the midpoint values. The same
qualitative improvement was observed in
Refs.~\cite{Barca:2024hrl,Barca:2026bmo} at heavier pion masses, as
expected when the variational basis suppresses current-enhanced
scalar-isoscalar contributions. Insertion-time dependence remains
visible, however, which suggests that a mild dependence on the
sink-source separation may not itself be sufficient as a signal
of ground-state dominance.

To examine the residual time dependence, we construct $R_{\rm LG}$ as
in \cref{eq:RLG}, using the gap determined from the standard ratio on
each ensemble and propagating its uncertainty. The comparison in
\cref{fig:a-ensembles-gevp-laplace} shows that filtering shifts the
A24 midpoints downward, most strongly at the shorter separations, and
exposes a separation dependence that is much less apparent before
filtering. This provides qualitative support for the competing
excited-state contributions seen on B64. The A48 filtered midpoint
central values show a similar tendency, but the filtering shifts are
less pronounced.

These comparisons support the observation that a nearly
separation-independent ratio after GEVP improvement can retain
excited-state contamination. The filtering response is qualitatively
similar to B64, particularly on A24, but does not resolve the residual
excitation energies precisely.

\section{ChPT Pion-Mass Corrections}
\label{app:pion-mass-corrections}

Following Refs.~\cite{Hoferichter:2015hva,Hoferichter:2023ptl}, we use
the ChPT expansion of $\sigma_{\pi N}$ through $\mathcal O(p^4)$ to
examine pion-mass corrections beyond the leading prescription of
\cref{sec:pion-mass-corrections}. We evaluate additive and
multiplicative corrections for the lattice results and compare the
phenomenological mass shift with Ref.~\cite{Hoferichter:2023ptl}.

At pion mass $M$, the $\mathcal O(p^2)$ and $\mathcal O(p^3)$ ChPT
expressions are~\cite{Hoferichter:2015hva,Hoferichter:2023ptl}
\begin{subequations}
\label{eq:sigma-chpt-orders}
\begin{align}
\sigma_{\pi N}^{(2)}(M)
 &= -4c_1M^2, \label{eq:sigma-chpt-p2}\\
\sigma_{\pi N}^{(3)}(M)
 &= \sigma_{\pi N}^{(2)}(M)
 -\frac{9g_A^2M^3}{64\pi F_\pi^2}.
 \label{eq:sigma-chpt-p3}
\end{align}
\end{subequations}
We evaluate these expressions directly at the initial and target pion masses, using the order-specific Roy--Steiner values~\cite{Hoferichter:2015hva}
\begin{align*}
c_1^{(2)}&=-0.74(2)~\mathrm{GeV}^{-1},\\
c_1^{(3)}&=-1.07(2)~\mathrm{GeV}^{-1}.
\end{align*}
We use $g_A=1.27641(56)$ and $F_\pi=92.215(143)$~MeV, as quoted by the
Particle Data Group
(PDG)~\cite{ParticleDataGroup:2024cfk,ParticleDataGroup:2010dbb}.

At $\mathcal O(p^4)$, the lower-order terms are evaluated with the
low-energy constants (LECs) appropriate to that order, and the
expansion becomes
\begin{subequations}
\label{eq:sigma-chpt-p4-ingredients}
\begin{align}
\sigma_{\pi N}^{(4)}(M)
 &= \sigma_{\pi N}^{(3)}(M)
 +M^4\left[\mathcal A(M)+2\mathcal B\right],
 \label{eq:sigma-chpt-p4}\\
\mathcal A(M)
 &= -\frac{3\left[g_A^2+m_N^{(0)}(-8c_1+c_2+4c_3)\right]}
 {64\pi^2F_\pi^2m_N^{(0)}}
 \notag\\
 &\quad\times\left[4\ln\!\left(\frac{M}{m_N^{(0)}}\right)+1\right],
 \label{eq:sigma-chpt-A}\\
\mathcal B
 &=e_1-\frac{3(2g_A^2-c_2m_N^{(0)})}
 {128\pi^2F_\pi^2m_N^{(0)}}
 +\frac{c_1(\bar\ell_3-1)}{16\pi^2F_\pi^2}.
 \label{eq:sigma-chpt-B}
\end{align}
\end{subequations}
We use $c_1=-1.11(3)$, $c_2=3.13(3)$, and $c_3=-5.61(6)$, in units of $\mathrm{GeV}^{-1}$~\cite{Hoferichter:2015hva}, with correlation matrix
\begin{align}
\rho(c_1,c_2,c_3)=
\begin{pmatrix}
1 & 0.18 & 0.58\\
0.18 & 1 & -0.64\\
0.58 & -0.64 & 1
\end{pmatrix}.
\label{eq:sigma-chpt-correlation}
\end{align}
We use $\bar\ell_3=3.41(41)$~\cite{Aoki:2013ldr}. We set the
chiral-limit nucleon mass to $m_N^{(0)}=869.6$~MeV by matching the
nucleon-mass expansion to $m_p=938.27$~MeV at
$M_{\pi^\pm}=139.57$~MeV. We determine $e_1$ through $\sigma_{\pi
  N}^{(4)}(M_{\pi^\pm})=59.0(3.5)$~MeV~\cite{Hoferichter:2023ptl},
obtaining $e_1=12(6)$~GeV$^{-3}$.

Let $\sigma_{\pi N}^\chi$ denote the ChPT expression at the chosen order. For $\sigma_{\pi N}^{\rm sim}$ normalized with the simulated bare light-quark mass, we define
\begin{align}
\sigma_{\pi N}^{\rm add}&=\sigma_{\pi N}^{\rm sim}+\Delta\sigma^\chi,
\label{eq:sigma-pion-additive}\\
\sigma_{\pi N}^{\rm mult}&=\sigma_{\pi N}^{\rm sim}r^\chi,
\label{eq:sigma-pion-multiplicative}
\end{align}
where
\begin{align}
\Delta\sigma^\chi(M_i\!\to M_f)
 &=\sigma_{\pi N}^\chi(M_f)-\sigma_{\pi N}^\chi(M_i),
\label{eq:sigma-pion-difference}
\end{align}
and
\begin{align}
r^\chi(M_i\!\to M_f)
 &=\frac{\sigma_{\pi N}^\chi(M_f)}{\sigma_{\pi N}^\chi(M_i)}.
\label{eq:sigma-pion-ratio}
\end{align}
We use $M_i=130.6(4)$~MeV for A48 and $140.2(2)$~MeV for B64~\cite{ExtendedTwistedMass:2022jpw}, with $M_f=m_\pi^\star$. The corrections in \cref{tab:sigma-chpt-corrections} show little change from $\mathcal O(p^3)$ to $\mathcal O(p^4)$.
\begin{table}[tbp]
\centering
\caption{Pion-mass corrections to the isoQCD target $m_\pi^\star=135.0$~MeV at successive ChPT orders. The last two columns give the additive shift $\Delta\sigma^\chi$ and multiplicative factor $r^\chi$ defined in \cref{eq:sigma-pion-difference,eq:sigma-pion-ratio}. The initial masses are $130.6(4)$~MeV for A48, $140.2(2)$~MeV for B64, and $M_{\pi^\pm}=139.57$~MeV for phenomenology (Phen.).}
\label{tab:sigma-chpt-corrections}
\centering
\begin{ruledtabular}
\begin{tabular}{cccc}
Initial point & Order & $\Delta\sigma^\chi$ [MeV] & $r^\chi$ \\ \hline
\multirow{3}{*}{A48}
 & $\mathcal O(p^2)$ & $+3.5(3)$ & $1.069(7)$ \\
 & $\mathcal O(p^3)$ & $+3.0(3)$ & $1.056(5)$ \\
 & $\mathcal O(p^4)$ & $+2.9(5)$ & $1.055(7)$ \\ \hline
\multirow{3}{*}{B64}
 & $\mathcal O(p^2)$ & $-4.2(2)$ & $0.927(3)$ \\
 & $\mathcal O(p^3)$ & $-3.6(2)$ & $0.941(2)$ \\
 & $\mathcal O(p^4)$ & $-3.5(5)$ & $0.942(6)$ \\ \hline
\multirow{3}{*}{Phen.}
 & $\mathcal O(p^2)$ & $-3.7(1)$ & $0.9356$ \\
 & $\mathcal O(p^3)$ & $-3.2(1)$ & $0.9475(3)$ \\
 & $\mathcal O(p^4)$ & $-3.0(5)$ & $0.948(5)$
\end{tabular}
\end{ruledtabular}
\end{table}
At $\mathcal O(p^2)$, $r^\chi=(M_f/M_i)^2$ reproduces the leading prescription in \cref{eq:light-mass-lo-correction}.

The phenomenology rows are consistent with the downward shifts $3.7(1)$, $3.2(1)$, and $3.1(1)(4)$~MeV at $\mathcal O(p^2)$, $\mathcal O(p^3)$, and $\mathcal O(p^4)$ in Ref.~\cite{Hoferichter:2023ptl}, whose target is the neutral-pion mass. Its first uncertainty comes from the LECs, and the second at $\mathcal O(p^4)$ comes from the phenomenological $\sigma_{\pi N}$ input used to determine $e_1$. The adopted phenomenological comparison is given in \cref{eq:sigma-phenomenology-isoqcd}.

\section{Laplace-Discretization Stability}
\label{app:laplace-discretization}

\begin{figure}[!htbp]
\centering
\includegraphics[width=\columnwidth]{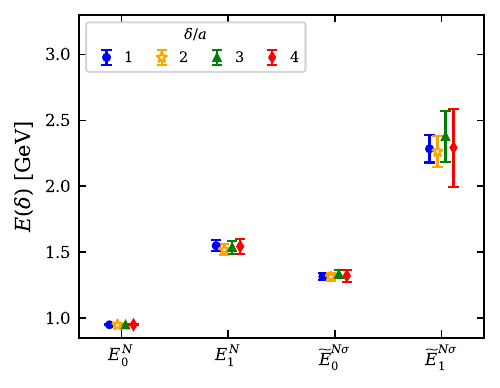}
\caption{B64 Laplace-filtered two-point energies. Colors and symbols distinguish the displacements, with open stars marking the main-analysis choice. The filter acts on the standard nucleon and GEVP-improved $N\sigma$ correlators.}
\label{fig:laplace-discretization-2pt}
\end{figure}

\begin{figure}[!htbp]
\centering
\includegraphics[width=\columnwidth]{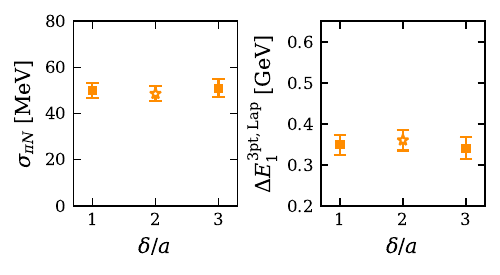}
\caption{B64 standard-ratio Laplace fits: $\sigma_{\pi N}$ (left) and $\Delta E_1^{\rm 3pt,Lap}$ (right) versus $\delta/a$. Open stars mark the main-analysis choice.}
\label{fig:laplace-discretization-3pt}
\end{figure}

The discrete filter in \cref{eq:laplace-filter} removes a contribution at the chosen energy for any displacement $\delta$, but its response to other energies and statistical fluctuations depends on $\delta$. We test this dependence on B64 while holding the original, unfiltered time ranges fixed. The energies in \cref{fig:laplace-discretization-2pt} remain compatible over $\delta/a=1,2,3,4$ for both channels.

The single-filter three-point results in \cref{fig:laplace-discretization-3pt} likewise show only mild changes in both central values and uncertainties over $\delta/a=1,2,3$.

The strange-current comparison in \cref{fig:sigma-s-laplace} uses the light-current gap $\Delta E_1^{\rm 3pt,Lap}$ with its uncertainty and $\delta/a=3$. The original insertion cut of $2a$ becomes $5a$ after filtering. The filtered data are too noisy to constrain the residual excited-state contamination.

\section{Approximating the Off-Diagonal Contractions}
\label{app:contraction-approximations}

\begin{figure}[!htbp]
\centering
\includegraphics[width=\columnwidth]{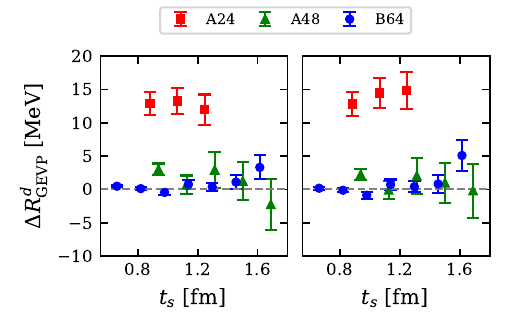}
\caption{Differences between complete and truncated light-current GEVP-improved midpoint ratios on A24, A48, and B64. Left: omit $B,W,Z$ from the off-diagonal entries. Right: retain only the two current-enhanced topologies and their sink counterparts in these entries. In both panels the truncated ratio is subtracted from the complete ratio, with correlated errors. Colors and symbols identify the ensembles. Gray dashed lines mark zero.}
\label{fig:contraction-approximations}
\end{figure}

We test whether simpler contractions can approximate the off-diagonal three-point functions. We compare two truncations of the topologies in \cref{fig:diags-3pt}: omitting the quark-connected $B,W,Z$ contributions, and retaining only the current-enhanced $N$-$J\sigma_i$ and $N$-$J$-$\sigma_i$ contributions together with their sink counterparts~\cite{Barca:2025det}. The latter includes both the connected current--sigma contraction and the separate current and sigma loops. In both cases $\braket{{\cal J}_N\mathcal O{\cal J}_N^\dagger}$, the eigenvectors, weight $W$, and two-point denominator are unchanged.

We assess each approximation through the difference from the complete $R_{\rm GEVP}^{d}$ ratio, keeping the correlations between the two determinations. The midpoint differences are shown in \cref{fig:contraction-approximations}. Although the absolute shifts are larger on A24, they amount to only $6$--$8\%$ of its much larger midpoint signal. On A48 and B64 the shifts are generally a few MeV or less, with larger uncertainties at the longest separations. Thus both truncations reproduce most of the midpoint signal. Their usefulness as less costly approximations depends on the target precision, since the omitted terms remain relevant for resolving effects at the level of these shifts.

\end{document}